\documentclass[journal,12pt]{IEEEtran}

\usepackage{cite}
\usepackage{url}
\usepackage{graphicx}
\usepackage{color}
\usepackage{float}
\usepackage{tabularx,colortbl}
\usepackage{ifthen}
\usepackage[greek,english]{babel}

\usepackage{amsmath}
\usepackage{amsthm,amsfonts}
\usepackage{amssymb}
\usepackage{pgfplots}

\usepackage{tikz}
\usepackage[customcolors]{hf-tikz}
\usepgfplotslibrary{colorbrewer}
\usetikzlibrary{mindmap,trees}
\usetikzlibrary{%
  3d,
  arrows,%
  shapes.misc,
  shapes.arrows,%
  chains,%
  calc,%
  matrix,%
	fit,%
  positioning,
  scopes,%
  decorations.pathmorphing,
  shadows,%
  decorations.markings,
  shapes.geometric,
  arrows.meta,bending,
	patterns,
	intersections, 
	pgfplots.fillbetween
}

\usepackage{tikz-3dplot}
\usepackage[pdfusetitle]{hyperref}
\hypersetup{
    colorlinks,
    linkcolor={red!50!black}, 
    citecolor={blue!50!black},
    urlcolor={blue!50!black}
}

\usepackage{cellspace}
\newcommand{\gpi}{\textrm{\greektext p}}
\renewcommand{\pi}{\gpi}
\providecommand*{\mat}[1]{\mathbf#1}
\providecommand*{\M}[1]{\mathbf#1}
\providecommand*{\tM}[1]{\tilde{\mathbf#1}}
\providecommand*{\mrm}[1]{\mathrm{#1}}

\renewcommand{\vec}[1]{{\boldsymbol#1}}
\providecommand*{\V}[1]{\boldsymbol#1}
\providecommand*{\UV}[1]{\hat{\boldsymbol#1}}
\providecommand*{\T}[1]{\mathrm{#1}}
\DeclareMathAccent{\ring}{\mathalpha}{operators}{"17}

\providecommand*{\eu}{\ensuremath{\mrm{e}}}

\providecommand*{\ju}{\ensuremath{\mrm{j}}}

\providecommand*{\diff}{\operatorname{d}\!}

\providecommand*{\diffS}{\operatorname{dS}\!}
\providecommand*{\diffV}{\operatorname{dV}\!}
\providecommand*{\svd}{\operatorname{svd}}
\providecommand*{\eig}{\operatorname{eig}}

\newcommand{\Tr}{\mathop{\mrm{Tr}}\nolimits}

\newcommand{\norm}[1]{\lVert#1\rVert}

\newcommand{\R}{\mathbb{R}{}}

\newcommand{\ie}{\textit{i.e.}\/, }
\newcommand{\eg}{\textit{e.g.}\/, }
\newcommand{\cf}{\textit{cf.}\/, }
\providecommand*{\acos}{\ensuremath{\mrm{acos}}}

\newcommand{\trans}{\text{T}}
\newcommand{\herm}{\text{H}}
\newcommand{\Id}{\mat{1}}

\newcommand{\maximize}{\mrm{maximize}}

\newcommand{\subto}{\mrm{subject\ to}}

\newcommand{\Aeff}{A_\mrm{eff}}
\newcommand{\As}{A_\mrm{s}}

\newcommand{\Na}{N_\mrm{A}} 
\newcommand{\Nk}{N_\mrm{c}} 
\newcommand{\reg}{\varOmega}

\newcommand{\regT}{\varOmega_\T{T}}
\newcommand{\sregT}{A_\T{T}}
\newcommand{\regR}{\varOmega_\T{R}}
\newcommand{\sregR}{A_\T{R}}
\newcommand{\sregTR}{A_\T{TR}}
\newcommand{\rv}{\vec{r}}
\newcommand{\rvh}{\UV{r}}

\newcommand{\Ts}[1]{}

\colorlet{dpurple}{blue!50!red}
\colorlet{dblue}{blue!50!black}
\colorlet{dgreen}{green!50!black}
\colorlet{dred}{red!50!black}
\colorlet{dyellow}{yellow!50!black}
\colorlet{dorange}{orange!50!black}
\definecolor{metal}{RGB}{218,165,32}
\definecolor{diel}{RGB}{1,165,32}
\definecolor{antenna}{RGB}{100,150,162}
\definecolor{breg}{rgb}{0.2,0.6,0.8}%
\definecolor{preg}{rgb}{0.8,0.2,0.2}%
\definecolor{reg}{RGB}{218,165,32}
\colorlet{treg}{blue!50!white}
\colorlet{rreg}{red!50!white}
\tikzset{>=latex}

\def\figw{0.55} 

\graphicspath{{figures/}{tikz/}{}}

\makeatother

\title{Shadow Area and Degrees of Freedom for Free-Space Communication}

\author{Mats Gustafsson
\thanks{Manuscript received August 6, 2024; revised \today. This work was supported by ELLIIT - an Excellence Center at Linkoping-Lund in Information Technology, the TICRA foundation, the Swedish Research Council SEE-6GIA, and SSF Sabbatical.}
\thanks{M. Gustafsson is with Lund University, Lund, Sweden, (e-mails: mats.gustafsson@eit.lth.se).}
}

\tdplotsetmaincoords{70}{140}
\pgfmathsetmacro{\wx}{0.8} 
\pgfmathsetmacro{\wy}{3.1} 
\pgfmathsetmacro{\wz}{0.4} 
\pgfmathsetmacro{\fx}{0} 
\pgfmathsetmacro{\fy}{1}
\pgfmathsetmacro{\fz}{0.4}
\pgfmathsetmacro{\fw}{0.2}
\pgfmathsetmacro{\hx}{2} 
\pgfmathsetmacro{\hy}{2.7}
\pgfmathsetmacro{\hz}{1}
\pgfmathsetmacro{\d}{0.2}
\tikzset{%
		grid/.style={very thin,gray},
		axis/.style={->,white,thin},
		cube/.style={fill=metal,fill opacity=0.5,draw=blue!50!black},
    horn/.style={fill=black!50!blue,fill opacity=0.5,draw=blue!50!black},
		cube hidden/.style={fill=black!50!blue,fill opacity=0.5,draw=blue!50!black},
    xyplane/.style={canvas is xy plane at z=#1,very thin}    
    }

\begin{document}

\onecolumn

\maketitle
\begin{abstract}
The number of degrees of freedom (NDoF) in a communication channel fundamentally limits the number of independent spatial modes available for transmitting and receiving information. Although the NDoF can be computed numerically for specific configurations using singular value decomposition (SVD) of the channel operator, this approach provides limited physical insight. In this paper, we introduce a simple analytical estimate for the NDoF between arbitrarily shaped transmitter and receiver regions in free space. In the electrically large limit, where the NDoF is high, it is well approximated by the mutual shadow area, measured in units of wavelength squared. This area corresponds to the projected overlap of the regions, integrated over all lines of sight, and captures their effective spatial coupling. The proposed estimate generalizes and unifies several previously established results, including those based on Weyl’s law, shadow area, and the paraxial approximation. We analyze several example configurations to illustrate the accuracy of the estimate and validate it through comparisons with numerical SVD computations of the propagation channel. The results provide both practical tools and physical insight for the design and analysis of high-capacity communication and sensing systems.
\end{abstract}


\section{Introduction}
Electromagnetic degrees of freedom (DoF) are critical parameters in communication systems, providing a fundamental estimate of the system's potential performance~\cite{Pizzo+etal2022,Hu+etal2018,Migliore2006a,Migliore2008,Franceschetti2017,Bjornson+etal2024,Poon+etal2005,Franceschetti+etal2009}. The number of degrees of freedom (NDoF) is particularly significant, as the capacity of a communication system tends to increase approximately linearly with the NDoF. This relationship underscores the importance of accurately determining and maximizing the NDoF in the design and analysis of advanced communication systems.

In the context of wave-based communication, DoF encompasses various components, including temporal, spatial, and polarization aspects. Each of these components contributes to the overall capacity and efficiency of the communication system~\cite{Franceschetti2017}. 
The concept of NDoF extends beyond communication systems and plays a crucial role in fields such as imaging, radar, and antennas. Early investigations by Di Francia in the mid-20th century highlighted the significance of NDoF in these areas~\cite{DiFrancia1955,DiFrancia1956}. For instance, in imaging, NDoF determines the resolution and quality of the reconstructed images~\cite{Bucci+Franceschetti1989,Bucci+Isernia1997,Pierri+Moretta2021,Maisto+etal2021}. 

Wave propagation in canonical geometries, such as spherical symmetric setups or rectangular waveguides, has traditionally been used to determine the NDoF for both electromagnetic and scalar fields~\cite{Bucci+Franceschetti1989,Bucci+Isernia1997,Kildal+etal2017,Hill1994,Janaswamy2011}. These findings can be interpreted as specific manifestations of Weyl's law~\cite{Weyl1911,Arendt+etal2009}, formulated in the early twentieth century by Hermann Weyl. Weyl's law describes the asymptotic eigenvalue distributions of Laplace and Helmholtz equations in bounded regions, demonstrating that the number of propagating waves, and hence NDoF, is proportional to the region's size, measured in wavelengths~\cite{Gustafsson2025a}.

These estimates are instrumental in determining the NDoF in arbitrarily shaped waveguides and for radiation emanating from convex-shaped regions. Additionally, Landau's eigenvalue problem offers a framework for analyzing NDoF~\cite{Franceschetti2017}. Numerical techniques, such as singular value decompositions (SVD) combined with free-space propagation models utilizing Green's function (dyadic), further enable the determination of NDoF for regions of arbitrary shapes~\cite{Piestun+Miller2000,Miller2019,Jensen+Wallace2008,Nordebo+etal2006}. In recent years, extensive research has focused on future communication systems, necessitating a deeper understanding of wave propagation and NDoF in complex environments~\cite{Pizzo+etal2022,Hu+etal2018,Ji+etal2023,Yuan+etal2024,Yuan+etal2022,Migliore2019,Ruiz-Sicilia2023}. 

In this work, we present a compact analytical estimate for the NDoF between arbitrary transmitter and receiver regions in free space. 
The key idea is that a high NDoF arises when the electrical size of the transmitter and receiver regions is large, \ie in the short-wavelength regime. In this asymptotic limit, the NDoF is well approximated by the mutual shadow area, \ie the projected overlap of the two regions integrated over all lines of sight, measured in units of wavelength squared.
This quantity captures the effective geometric coupling between the regions and serves as a unifying metric that incorporates and generalizes prior results from Weyl's law, shadow area~\cite{Gustafsson2025a}, and the paraxial approximation~\cite{Miller2019}.

We validate this estimate using numerical simulations of the channel operator's singular values for a range of 2D and 3D setups. We show that the number of near-unity singular values closely matches the proposed analytical NDoF, $\Na$, and that normalized eigenvalue distributions tend to cluster for systems with equal $\Na$, regardless of specific geometry. To handle large-scale problems efficiently, we also employ a randomized SVD algorithm that leverages the low-rank structure of the propagation operator~\cite{Halko+etal2011}.

The remainder of this paper is organized as follows. Section~\ref{S:FreeSpaceCom} introduces the free-space channel model. Section~\ref{S:NDoFshadowArea} describes the NDoF estimate and mutual shadow area for fully enclosing receiver structures. Section~\ref{S:PartlyCircumscribing} extends the analysis to partially enclosing receivers. Section~\ref{S:CommVolMutualShadowArea} discusses communication between volumes. Numerical results in 2D and 3D are presented in Secs~\ref{S:2D} and~\ref{S:3D}, respectively. The paper is concluded in Sec.~\ref{S:Conclusions} and derivations and technical details are provided in the appendices.


\section{Free-space communication channel}\label{S:FreeSpaceCom}
Communication between antenna systems confined to a transmitting region $\regT$ and a receiving region $\regR$ can be modeled by the radiation from current densities $\V{J}_\T{T}(\V{r})$ in $\regT$ to fields $\V{E}_\T{R}(\rv)$ in $\regR$, as illustrated in Fig.~\ref{fig:TRregsH}. For simplicity, we consider transmitting electric currents $\V{J}_\T{T}$ and receiving electric fields $\V{E}_\T{R}$, but the results and model are also applicable to scenarios involving both electric and magnetic currents, as well as electric and magnetic fields.

\newcommand{\Treg}{(-1,-1.3) (0,-1.2) (0.5,-0.1) (2,0) (1,0.7) (0,1) (-0.5,0.5)}
\newcommand{\Tregc}{(-1,-1.3), (0,-1.2), (0.5,-0.1), (2,0), (1,0.7), (0,1), (-0.5,0.5)}
\newcommand{\Rreg}{(-1,-1.3) (0,-1.2) (0.5,-0.5) (2,0) (1.5,0.8) (1,0.7) (0,1) (-0.5,0.5)}
\newcommand{\Rregc}{(-1,-1.3), (0,-1.2), (0.5,-0.5), (2,0), (1.5,0.8), (1,0.7), (0,1), (-0.5,0.5)}

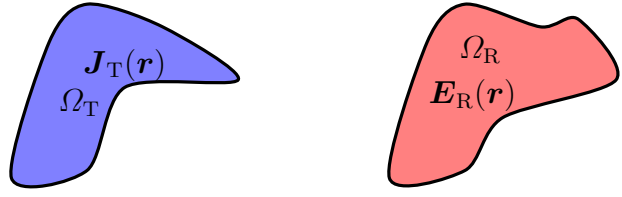
\begin{figure}[t]%
{\centering
\begin{tikzpicture}[scale=1,>= stealth]
	\draw [fill=treg,very thick] plot [smooth cycle,scale=1] coordinates {\Treg};    
	\node at (-0.1,-0.3) {$\reg_\T{T}$};	
	\node at (0.5,0.2) {$\vec{J}_\T{T}(\rv)$};	
\begin{scope}[xshift=5cm]
	\draw [fill=rreg,very thick,rotate=0] plot [smooth cycle,scale=1] coordinates {\Rreg};    
	\node at (0.1,-0.2) {$\vec{E}_\T{R}(\rv)$};	
	\node at (0.2,0.4) {$\reg_\T{R}$};	
\end{scope}	
\end{tikzpicture}
\vspace{-1mm}
\par}
\caption{Communication channel between a transmitting region $\reg_\T{T}$ and receiving region $\reg_\T{R}$, represented by the radiated field $\V{E}_\T{R}$ in $\regR$ generated by current densities $\V{J}_\T{T}$ in $\regT$.}%
\label{fig:TRregsH}%
\end{figure}

The transfer of information from $\regT$ to $\regR$ depends on the number of independent temporal and spatial channels that can be formed between these regions~\cite{Franceschetti2017}. The number of temporal channels is proportional to the available bandwidth~\cite{Paulraj+etal2003,Molisch2011} and is not strongly linked to the geometrical configuration of the regions for electrically large structures. Here, we focus on the spectral efficiency (capacity per unit bandwidth)~\cite{Paulraj+etal2003,Molisch2011}. Spectral efficiency depends on the spatial channels between the regions and the signal-to-noise ratio (SNR).

The spectral efficiency, or capacity, is determined by Shannon's formula for a channel formed by antenna systems in $\regT$ and $\regR$ at a given SNR~\cite{Paulraj+etal2003,Molisch2011}. This capacity increases with an improved SNR and by including more (ideal) antenna elements in the regions. The increase in capacity is nearly linear up to a certain number of elements and saturates beyond this point~\cite{Franceschetti2017}. Increasing the number of elements after saturation requires a significantly higher SNR to further enhance the capacity. This threshold number of antenna elements (ports) is referred to as the number of degrees of freedom (NDoF)~\cite{Franceschetti2017}. The NDoF depends on the geometrical configuration of $\regT$ and $\regR$ and the frequency $f$ or wavelength $\lambda = \T{c}/f$, where $\T{c}$ denotes the speed of light.

The NDoF between two regions is often determined from the channel formed by a transmitting array in $\reg_{\T{T}}$ and a receiving array in $\reg_{\T{R}}$. These arrays are created by discretizing the regions into sub-wavelength spaced point or dipole elements~\cite{Piestun+Miller2000,Miller2019}. This approach is analogous to decomposing the current density into a set of basis functions, where each basis function corresponds to an array element~\cite{Harrington1968}. Using $N_{\T{T}}$ transmitting elements and $N_{\T{R}}$ receiving elements defines a channel matrix $\M{H}$ with dimensions $N_{\T{R}} \times N_{\T{T}}$.

\begin{figure}[t]
{\centering
\begin{tikzpicture}[scale=1,>= stealth]
\foreach \coord [count=\i] in \Tregc {
    \coordinate [at=\coord, name=A\i];
    \foreach \coord [count=\j] in \Rregc {
        \coordinate [at=\coord, name=B\j];
        \draw[green!50!white,very thin] (A\i) -- ($(B\j)+(5,0)$);
}}
	\draw [fill=treg,very thick,mark=o] plot [smooth cycle,scale=1] coordinates {\Treg};    
	\node at (-0.1,-0.3) {$\reg_\T{T}$};	
\begin{scope}[xshift=5cm]
	\draw [fill=rreg,very thick,rotate=0,mark=o] plot [smooth cycle,scale=1] coordinates {\Rreg};    
	\node at (0.2,0.4) {$\reg_\T{R}$};	
\end{scope}

\end{tikzpicture}
\vspace{-1mm}
\par}
\caption{Communication channel represented as paths between transmitting and receiving array elements, here represented by 7 transmitters and 8 receivers indicated by the markers.}%
\label{fig:TXchannel}%
\end{figure}
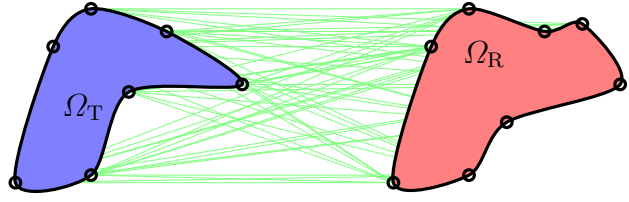

Forming an SVD of the channel matrix $\M{H}$ reveals a threshold number, after which the singular values decrease rapidly. The slope of this decay increases with increasing frequency (decreasing wavelength). 
The effective NDoF~\cite{Shiu+etal2000,Yuan+etal2022} and effective rank~\cite{Roy+Vetterli2007} are ways to approximate this threshold with  the effective NDoF given by 
\begin{equation}
	N_{\mrm{e}} 
	=
\frac{\norm{\M{H}}_\T{F}^4}{\norm{\M{H}\M{H}^{\herm}}_\T{F}^2} 
	= \frac{\left(\sum_{n}\sigma_n\right)^2}{\sum_{n}\sigma_n^2},
 \label{eq:eNDoF}
\end{equation}
where $\sigma_n$ denotes the squared singular values of $\M{H}$, \ie eigenvalues of the correlation matrix 
\begin{equation}
    \sigma_n=\eig(\M{H}\M{H}^{\herm}), 
    \label{eq:eigHH}
\end{equation}
the superscript ${}^{\herm}$ denotes Hermitian transpose, and $\norm{\cdot}_\T{F}$ is the Frobenius norm~\cite{Horn+Johnson1991}. 
These approaches and approximations have been thoroughly investigated in the literature~\cite{Shiu+etal2000,Yuan+etal2022}. The effective NDoF is easy to evaluate and useful for arbitrary configurations and wavelengths.

The effective NDoF and distribution of the singular values are independent of the normalization of the channel matrix $\M{H}$. For the presentation in this paper, it is convenient to normalize the channel matrix with its Frobenius norm $\norm{\M{H}}_\T{F}$, \ie $\widetilde{\M{H}}= \M{H}/\norm{\M{H}}_\T{F}$.  This normalization is equivalent to normalizing the eigenvalues $\sigma_n$ as 
\begin{equation}
    \zeta_n
    =\sigma_n/\sum_n\sigma_n
    \label{eq:HHeigNorm}
\end{equation}
with $\zeta_n$ denoting the eigenvalues of the normalized correlation matrix, \ie $\zeta_n=\eig(\widetilde{\M{H}}\widetilde{\M{H}}^{\herm})$. These normalized eigenvalues are positive $\zeta_n\geq 0$ and satisfy
\begin{equation}
    \sum \zeta_n=1
    \quad\text{and }
    \sum \zeta_n^2 = N_\T{e}^{-1}.
\end{equation}

Channels and channel eigenvalues $\zeta_n$~\eqref{eq:HHeigNorm} can be evaluated numerically~\cite{Miller2019} by sampling the channel, as illustrated in Fig.~\ref{fig:TXchannel}. Point sources, expansions in local basis functions, and global modes can be used to represent the channel. In this paper, we use point sources sampled with five points per wavelength together with the free-space Green's function in the numerical examples.

\begin{figure}
    \centering
    \includegraphics[width=\figw\columnwidth]{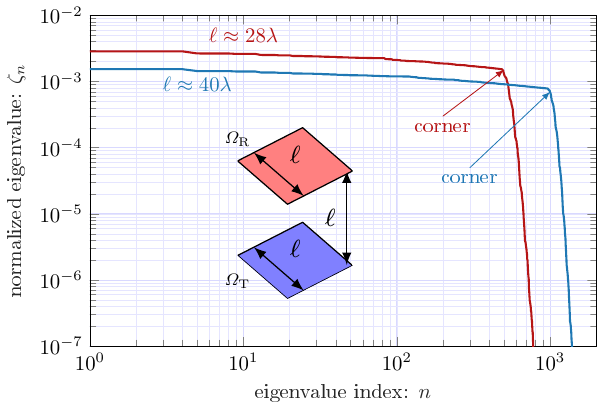}
    \caption{Distribution of the normalized eigenvalues, $\zeta_n$, in~\eqref{eq:HHeigNorm} for the free-space channel between two square regions at $\ell/\lambda\in\{28,40\}$, each with side length $\ell$, separated by a distance $\ell$, as depicted in the inset.}
    \label{fig:idealchannelNDoF}
\end{figure}

As an example, the normalized eigenvalues $\zeta_n$ of the free-space channel between two square plates are depicted in Fig.~\ref{fig:idealchannelNDoF} for $\ell/\lambda \in \{28, 40\}$ wavelengths per side. The eigenvalues remain relatively constant up to a \textit{corner} or \textit{knee}, after which they decay rapidly~\cite{Bucci+Franceschetti1989,Franceschetti2017,Miller2019,Solimene+etal2018,Solimene+etal2019,Bucci+Migliore2025,Maisto+etal2021,Kuang+etal2025}. The two presented cases have corners around 500 and 1000. These values also approximately match the results obtained using the effective NDoF~\eqref{eq:eNDoF}, based on the simulated eigenvalues.  

The main results of this paper are semi-analytical expressions for the corner of the eigenvalues, solely expressed in terms of the geometries $\regT$ and $\regR$. We focus on regions $\regT$ and $\regR$ such that there is a well-defined corner. For general cases, such as when multiple receivers are located at different distances from the transmitter, the eigenvalue spectrum can exhibit multiple corners~\cite{Solimene+etal2013}. In those cases, the presented results are valid for the last corner.

\section{Background on NDoF estimates in free space}\label{S:NDoFshadowArea}
\begin{figure}%
{\centering
\begin{tikzpicture}[scale=0.7,>= stealth]
	\begin{scope}
        \draw[fill=treg] (0,0) circle(1cm); 
	\node at (0,-0.4) {$\regT$};	
	\node at (0.0,0.2) {$\vec{J}_\T{T}(\rv)$};	
\draw[->,thick,decorate,decoration=snake] (-1,0) -- node[above] {$\M{f}$} (-1.7,1);	
\draw[dashed,rreg,very thick] (0,0) circle(2cm);
	\node at (-1.5,1.5) {$\regR$};	
	\end{scope}	
	\begin{scope}[xshift=5cm,scale=1]
	\draw [fill=treg,very thick] plot [smooth cycle,scale=0.85] coordinates {(-1,-1) (0,-1.2) (0.5,-0.1) (2,0) (1,0.9) (0,1) (-0.5,0.5)};    
	\node at (-0.1,-0.3) {$\regT$};	
	\node at (0.5,0.2) {$\vec{J}_\T{T}(\rv)$};	
\draw[->,thick,decorate,decoration=snake] (-1,0) -- node[below] {$\M{f}$} (-1.7,1);	
\draw[dashed,rreg,very thick] (0,0) circle(2cm);
	\node at (-1.5,1.5) {$\regR$};	
	\end{scope}	
 \begin{scope}[xshift=8cm,yshift=-2cm,scale=1.5]
\begin{axis}[%
width=1.141cm,
height=2.452cm,
at={(0,0)},
scale only axis,
plot box ratio=1 1 2,
xmin=-0.5,
xmax=0.5,
tick align=outside,
ymin=-0.5,
ymax=0.5,
zmin=0,
zmax=2,
view={-37.5}{30},
axis line style={draw=none},
ticks=none,
title style={font=\bfseries},
axis x line*=bottom,
axis y line*=left,
axis z line*=left
]

\addplot3[area legend, draw=black, fill=treg, forget plot]
table[row sep=crcr] {%
x	y	z\\
-0.5	-0.5	0\\
0.5	-0.5	0\\
0.5	0.5	0\\
-0.5	0.5	0\\
}--cycle;

\addplot3[area legend, draw=black, fill=rreg, forget plot]
table[row sep=crcr] {%
x	y	z\\
-0.5	-0.5	2\\
0.5	-0.5	2\\
0.5	0.5	2\\
-0.5	0.5	2\\
}--cycle;

\draw[<->] (axis cs: 0.45,-0.45,0) -- node[left] {$d$} (axis cs: 0.45,-0.45,2);
\node at (axis cs: 0.05,-0.05,2) {$\regR$};
\node at (axis cs: 0.05,-0.05,0) {$\regT$};
\end{axis}  
 \end{scope} 
\node at (-2,2) {a)};	
\node at (3,2) {b)};	
\node at (8,2) {c)};	
\end{tikzpicture}
\par}
\caption{Free-space channel and DoF for: a) regions with spherical  (cylindrical) symmetry, b) arbitrary shaped transmitter region $\regT$ circumscribed by a spherical (cylindrical) receiver region, c) two parallel regions separated by a distance $d$.}%
\label{fig:DoFsetups}%
\end{figure}
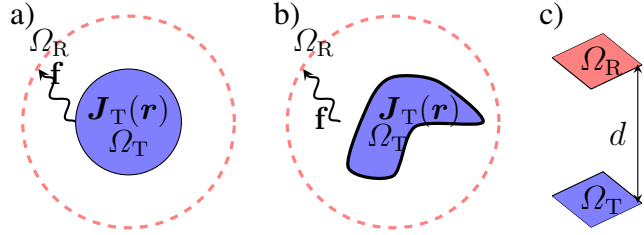

Analytic solutions of the wave equation or Maxwell's equations can be used to estimate the NDoF for canonical geometries, such as regions with spherical or cylindrical symmetries, by expanding the field in spherical and cylindrical waves, respectively~\cite{Stratton1941,Kristensson2016}.
The NDoF for a spherical region with radius $a$ surrounded by a larger spherical region, see Fig.~\ref{fig:DoFsetups}a, has been investigated using various techniques~\cite{Gustafsson+Nordebo2007a,Bucci+Isernia1997,Kildal+etal2017,Gustafsson+Lundgren2024,Harrington1958,Kuang+etal2025}. These studies have found that electrically small regions, with $ka \ll 1$, exhibit six (polarization) DoF, while regions with $ka \gg 1$ have approximately $(ka)^2$ DoF per polarization, where $k=2\pi/\lambda$ denotes the wavenumber. In the corresponding 2D case with circular (cylindrical) symmetry, the expansion in cylindrical waves yields approximately $2ka$ DoF for $ka\gg 1$. When expressed in terms of surface area $A$ and circumference $L$. The NDoFs for cylindrical and spherical regions become $2L/\lambda$ and $\pi A/\lambda^2$, respectively.

The NDoF for arbitrarily shaped transmitting regions $\regT$, see Fig.~\ref{fig:DoFsetups}b, is determined by investigating the ability of the transmitter to generate waves in directions $\UV{k}$ or, by reciprocity, to receive plane waves propagating in the opposite direction~\cite{Gustafsson2025a}. This concept leads to the notion of shadow area in the asymptotic limit of many NDoFs~\cite{Gustafsson+Lundgren2024,Gustafsson2025a}. The shadow (or projected) area of a region $\regT$ for a direction $\UV{k}$ quantifies how transmitters confined to the region can direct radiation in that direction. For convex shapes, this reduces to the classical cross-sectional area, which is commonly used to estimate the effective area of aperture antennas~\cite{Stutzman+Thiele1998}.

An object surrounded by a (convex) receiving region $\regR$, as in Fig.~\ref{fig:DoFsetups}b, can interact with plane waves propagating in any direction, necessitating a summation over all directions. The total shadow (or projected) area $\As = \int_{4\pi} \As(\UV{k}) \diff\Omega$ is defined as the shadow area $\As(\UV{k})$ integrated over the unit sphere (or circle in 2D), \ie over all directions. To simplify notation, we include the argument $\As(\UV{k})$ to denote the shadow area in a specific direction $\UV{k}$ and use $\As$ without argument for the total shadow area throughout this paper. The resulting asymptotic NDoF for an arbitrarily shaped object surrounded by a large spherical region is approximately~\cite{Gustafsson2025a}
\begin{equation}
    N_\T{A} = \frac{L_\T{s}}{\lambda}
    \overset{\T{convex}}{=}
    \frac{2L}{\lambda}
    \quad\text{and }
    N_\T{A} = \frac{\As}{\lambda^2}
    \overset{\T{convex}}{=}
    \frac{\pi A}{\lambda^2},
    \label{eq:NDoFshadowArea}
\end{equation}
where $\As$ denotes the total shadow area ($L_\T{s}$ for length) of the object and the wavelength $\lambda$ is much smaller than the object's dimensions. The total shadow area (and length) in~\eqref{eq:NDoFshadowArea} are proportional to the surface area $A$ (or the circumference $L$) for convex shapes~\cite{Vouk1948,Gustafsson2025a}. These results are consistent with Weyl's law for convex shapes~\cite{Gustafsson2025a}. Notably, the surface case differs from the product of two line cases by a factor of $\pi/4$, as also observed for planar rectangles~\cite{Pizzo+etal2022,Hu+etal2018}.

Note that the shadow area of a region does not correspond to a specific material but can be interpreted as the shadow cast by an opaque object filling the region. Here, it is used to quantify what can be constructed within the region. A specific realization, such as dipole arrays or dielectric antennas, can naturally have a smaller shadow area.

\begin{figure}
    \centering
    \begin{tikzpicture}
        \node at (0,0) {\includegraphics[width=0.45\columnwidth]{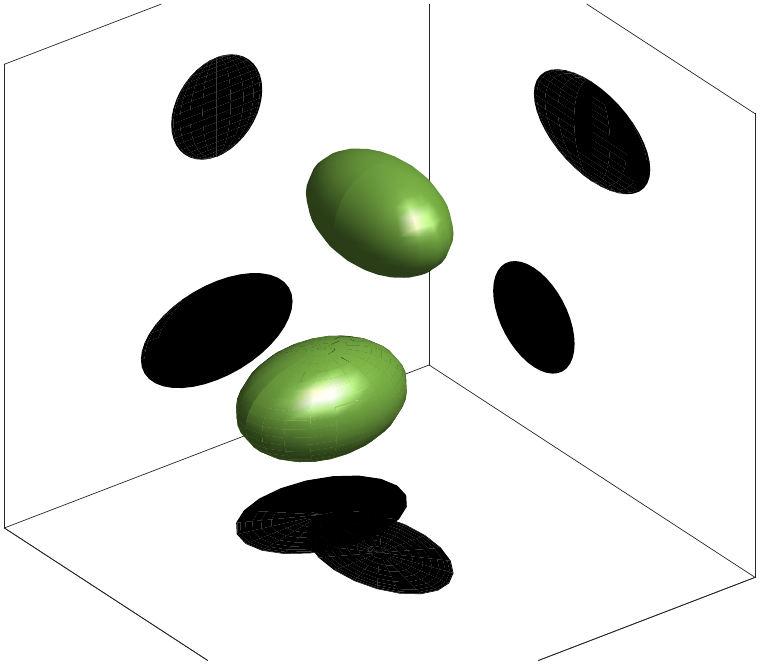}};
        \node[black] at (2.2,1) {$A_\T{s}(\UV{x})$};
        \node[black] at (-2,1.3) {$A_\T{s}(\UV{y})$};
        \node[white] at (-0.9,-2) {$A_\T{s}(\UV{z})$};
    \end{tikzpicture}
    \caption{Illustration of the shadow (projected) area $A_\T{s}(\UV{k})=A_\T{s}(-\UV{k})$ of two spheroids illuminated from the directions $\UV{x}$, $\UV{y}$, and $\UV{z}$ projected onto planes perpendicular to each respective illumination direction. }
    \label{fig:ShadowAreaXYZ}
\end{figure}
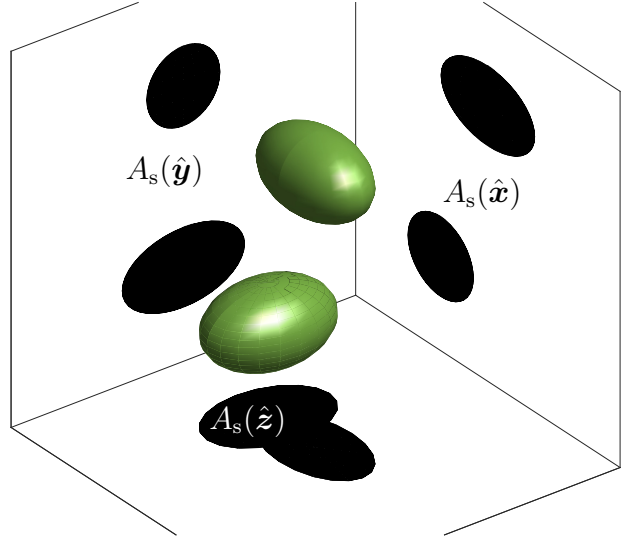

Shadow areas are illustrated in Fig.~\ref{fig:ShadowAreaXYZ} for an object composed of two separated spheroids illuminated in the $\UV{k} \in \{\UV{x},\UV{y},\UV{z}\}$ directions, where the shadow is projected onto a plane perpendicular to $\UV{k}$. The shadows from the $\{\UV{x},\UV{y}\}$ directions produce non-overlapping elliptical shadows, whereas illumination from $\UV{k} = \UV{z}$ results in a single combined shadow. The total shadow area $\As$ is determined by integrating the shadow area over all illumination directions. In this paper, we show that the overlapping shadow of the regions, measured in wavelengths, is proportional to the NDoF between the two regions.

\begin{figure}
    \centering
    \includegraphics[width=0.5\linewidth]{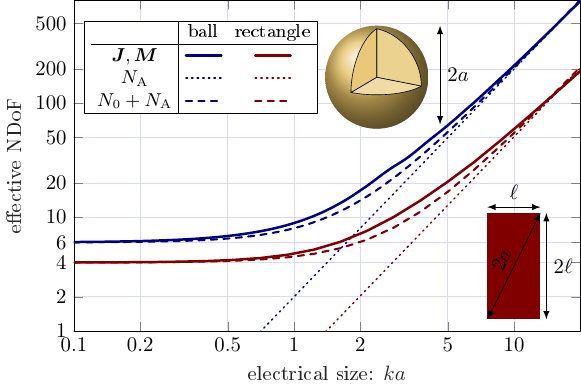}
    \caption{Effective NDoF based on radiation modes~\cite{Gustafsson2025a} for a ball and planar rectangle as a function of the electrical size $ka$ (solid curves) compared with asymptotic expression~\eqref{eq:NDoFshadowArea} (dotted) and linear combination with small size asymptotic $N_0+2A_\T{s}/\lambda^2$ (dashed) considering two polarizations. Here, $N_0$ denotes the small size asymptotic with $N_0=6$ for the ball and $N_0=4$ for the plate.}
    \label{fig:NDoFSphRec}
\end{figure}

The asymptotic NDoF $\Na$ represents the dominant term of the NDoF as $\Na$ becomes large. The NDoF increases with frequency (or decreases with wavelength), providing an intuitive interpretation of the asymptotic NDoF as a high-frequency limit. Lower-order terms, \eg those proportional to $\lambda^{-1}$, are neglected in these expansions. This approach is consistent with focusing on the dominant term in Weyl's law~\cite{Weyl1911,Arendt+etal2009} and, for example, neglecting the linear term in the NDoF of a spherical structure~\cite{Bucci+Isernia1997}, given by $N\approx 2L(L+2)\approx 2L^2$ with $L=ka$. Assuming these asymptotic expansions, the relative error is proportional to $\Na^{-1/2}$.  

The asymptotic NDoF is also illustrated in Fig.~\ref{fig:NDoFSphRec} for a sphere and a rectangular region. Here, the dotted lines correspond to the asymptotic NDoF~\eqref{eq:NDoFshadowArea}, the solid lines represent the effective NDoF~\eqref{eq:eNDoF} determined numerically, and the dashed curves show a linear combination of the low- and high-frequency asymptotics. The low-frequency asymptotic originates from the three electric and three magnetic dipoles for the sphere~\cite{Kristensson2016}, which reduce to four for the planar rectangle. For these cases, the asymptotic NDoF approximates the effective NDoF well for $ka \gtrsim 10$, while the linear combination provides a good approximation across all sizes.  

The total shadow area estimates the NDoF that can be transmitted from a region $\regT$ to the far field~\cite{Gustafsson2025a}. This provides a limit that is independent of the receiver location and configuration. For line-of-sight communication between well-separated planar regions, the paraxial approximation can be used~\cite{Piestun+Miller2000,Miller2019}. This asymptotic solution for narrow angular propagation provide closed form expressions for the NDoF for lines with lengths $L_\T{T}$ and $L_\T{R}$ and plates with areas $A_\T{T}$ and $A_\T{R}$ separated a distance $d$, see Fig.~\ref{fig:DoFsetups}c, as~\cite{Piestun+Miller2000,Miller2019}
\begin{equation}    
    N_\T{P} = \frac{L_\T{T}L_\T{R}}{d\lambda}
    \quad\text{and }
    N_\T{P} = \frac{A_\T{T}A_\T{R}}{d^2\lambda^2}
    \label{eq:NDoFparaxial}
\end{equation}
(per polarization) with $d$ much larger than the object sizes. 
Here, we note that the paraxial approximation for surfaces is a product of two line cases in contrast to~\eqref{eq:NDoFshadowArea}.

\section{Partly surrounding structure}\label{S:PartlyCircumscribing}

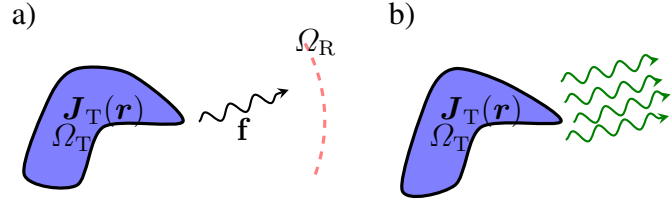
\begin{figure}%
{\centering
\begin{tikzpicture}[scale=1,>= stealth]
	\begin{scope}[xshift=0cm,scale=0.7]
	\draw [fill=treg,very thick] plot [smooth cycle,scale=1] coordinates {(-1,-1) (0,-1.2) (0.5,-0.1) (2,0) (1,0.9) (0,1) (-0.5,0.5)};    
	\node at (-0.1,-0.3) {$\regT$};	
	\node at (0.5,0.2) {$\vec{J}_\T{T}(\rv)$};	
\draw[->,thick,decorate,decoration=snake] (2.3,0) -- node[below] {$\M{f}$} (4,0.6);	
\draw[dashed,rreg,very thick] (4.5cm,-1cm) arc(-20:30:3cm);
	\node at (4.5,1.5) {$\regR$};	
	\node at (-1,2) {a)};	
	\end{scope}	
 \begin{scope}[xshift=5cm,yshift=0cm,scale=0.7]
 	\draw [fill=treg,very thick] plot [smooth cycle,scale=1] coordinates {\Treg};    
	\node at (-0.1,-0.3) {$\reg_\T{T}$};	
	\node at (0.5,0.2) {$\vec{J}_\T{T}(\rv)$};		 
\begin{scope}[rotate=10,xshift=5mm,yshift=-2mm]    
\draw[->,thick,decorate,decoration=snake,dgreen] (1.5,-0.5) --  (3.5,-0.4);	
\draw[->,thick,decorate,decoration=snake,dgreen] (1.7,-0.2) --  (3.6,-0.0);	
\draw[->,thick,decorate,decoration=snake,dgreen] (1.6,0.2) --  (3.5,0.3);	
\draw[->,thick,decorate,decoration=snake,dgreen] (1.6,0.6) --  (3.5,0.7);	
\end{scope}
\node at (-1,2) {b)};	
 \end{scope}
\end{tikzpicture}
\vspace{-1mm}
\par}
\caption{a) Communication from a transmitter in $\regT$ to a partly surrounding region $\regR$, \ie $\regR$ is a subset of the unit sphere representing directions $\UV{k}$. b) Representation
using rays propagating in directions $\UV{k}$.}%
\label{fig:RDoF}%
\end{figure}


NDoF for communication between transmitters in the region $\regT$ to receivers in the far field partly surrounding the transmitter, or similarly, on a large partly surrounding spherical sector $\regR$, as shown in Fig.~\ref{fig:RDoF}a is investigated first. It reduces to the case with surrounding receivers~\eqref{eq:NDoFshadowArea} in one limit and the paraxial approximation~\eqref{eq:NDoFparaxial} in the other. Current densities $\V{J}_\T{T}(\rv)$ in the transmitting region $\regT$ constitute sources for the radiated field and are expanded in a sufficiently large number of basis functions, see App.~\ref{S:currentdensity}. It is assumed that the transmitting region $\regT$ is constructed of some material with minimal material losses described by a resistivity~\cite{Gustafsson2025a}.
Similarly, the radiated field over the receiver region is obtained by evaluating the field over $\regR$. 

The capacity (or spectral efficiency) is formulated as an optimization problem over the current density $\V{J}_\T{T}$ in App.~\ref{S:CapRadmodes}. Diagonalization of the optimization problem introduces a decomposition of the current density in generalized radiating modes characterized by their efficiencies $\nu_n\geq 0$. Geometry and material properties of $\regT$ and configuration of $\regR$ are contained in these generalized radiation modes defined in App.~\ref{S:CapRadmodes}.
The capacity is solved using waterfilling~\cite{Paulraj+etal2003} over radiation mode efficiencies $\nu_n$, resulting in
\begin{equation}
	\max_{\sum \tilde{P}_n=1}\sum_{n=1}^{N} \log_2\big(1+\gamma \nu_n\tilde{P}_n\big),
\label{eq:CapRadm}
\end{equation}  
with a finite number of non-zero power levels $\tilde{P}_n$ associated with generalized radiation modes of sufficiently high efficiency. 
Water-filling algorithm solves~\eqref{eq:CapRadm} and distributes the power $\tilde{P}_n$ over the modes~\cite{Paulraj+etal2003} analogous to filling a bucket with heights proportional to $\nu_n^{-1}$ with water. The generalized radiation mode efficiencies $\nu_n$ in~\eqref{eq:CapRadm} correspond to the channel eigenvalues $\sigma_n$ in~\eqref{eq:eigHH} for a channel modeled with orthogonal currents. Normalizing according to~\eqref{eq:HHeigNorm} establishes a direct link with $\zeta_n$, as used in the numerical examples.

\begin{figure}
    \centering
    \includegraphics[width=\figw\columnwidth]{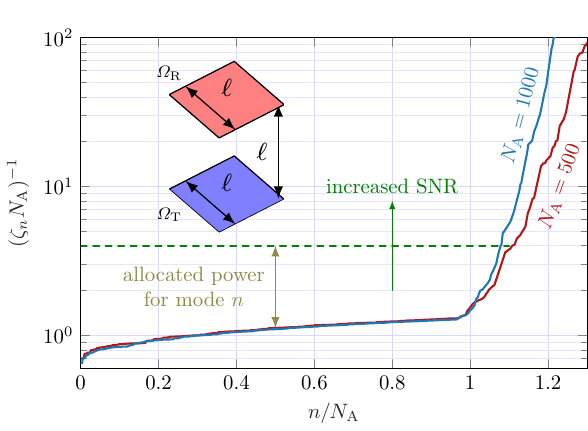}
    \caption{Reciprocal of the eigenvalues for the channel correlation matrix of two parallel squares with side length $\ell$ separated a distance $\ell$ using $\Na\in\{500,1000\}$, \cf Fig.~\ref{fig:idealchannelNDoF}. The modal index $n$ is normalized by $\Na$.}
    \label{fig:NDoFwaterfilling}
\end{figure}

The reciprocal of the normalized eigenvalues $(\zeta_n\Na)^{-1}$ are plotted in Fig.~\ref{fig:NDoFwaterfilling} for the channel between the two rectangles shown in Fig.~\ref{fig:idealchannelNDoF}. 
For a given SNR (power level), illustrated by the dashed line in Fig.~\ref{fig:NDoFwaterfilling} the allocated power is proportional to the distance down to $(\zeta_n\Na)^{-1}$. This illustrates the high cost in SNR to use more than $\Na$ modes. 

Determination of the NDoF from the number of generalized radiation modes~\eqref{eq:RadmEff} is easily achieved numerically for arbitrarily shaped objects, material parameters, and receiver configurations. Analytical solutions in the asymptotic limit $\Na\gg 1$, presented here, complement the numerical results and provide physical insight and understanding. 

\color{black}

The analytical solution, see App.~\ref{S:trace}, is determined by first relating the sum of the efficiencies $\nu_n$ to the maximum gain $G$ and effective area $A_{\T{eff}}$ of antennas designed within $\regT$
\begin{equation}
\sum_{n=1}^N \nu_n 
=
\frac{1}{4\pi}\sum_{\UV{e}=\{\UV{\theta},\UV{\phi}\}}\int_{\regR} G(\UV{k},\UV{e})\diff\Omega_{\UV{k}}
= \frac{1}{\lambda^2}\sum_{\UV{e}=\{\UV{\theta},\UV{\phi}\}}\int_{\regR} A_{\T{eff}}(\UV{k},\UV{e})\diff\Omega_{\UV{k}}.
\label{eq:NDoFMaxAeffInt}
\end{equation}
This identity generalizes the result in~\cite{Gustafsson2025a}, which considers receivers distributed over a full spherical region $\regR = 4\pi$, to the case of receivers that partially surround the transmitter, specifically, those covering an angular sector (see Fig.~\ref{fig:RDoF}a). The relation links port-observed powers, transmitted power, and received power: it characterizes the efficiency with which transmitting antennas in $\regT$ radiate power into $\regR$ as seen from the ports, the gain of antennas in $\regT$ integrated over the receiving region $\regR$, and the effective area of antennas in $\regT$ for receiving fields transmitted from $\regR$.

The generalized radiation mode efficiencies $\nu_n$ are further related to the geometrical structure by observing that the maximal effective area in a direction $\UV{k}$ approaches the geometrical cross section, $\As(\UV{k})$, in the electrically large limit~\cite{Gustafsson+Capek2019} $	\max\Aeff(\UV{k}) \approx \As(\UV{k})$. This connects the generalized radiation modes and geometrical properties of the region $\regT$ 
with the maximal effective area expressed solely in geometrical parameters producing the asymptotic relation
\begin{equation}
    \sum_{n=1}^N\nu_n \approx 
    \frac{2}{\lambda^2}\int_{\regR} \As(\UV{k})\diff\Omega_{\UV{k}}
    =\frac{2 A_\T{TR}}{\lambda^2},
    \label{eq:sumnu}
\end{equation}
where the factor of two stems from the two polarizations $\{\UV{\theta},\UV{\phi}\}$. The identity~\eqref{eq:sumnu} relates the efficiency of the generalized radiation modes with the shadow area $A_\T{TR}=\int_{\regR}\As(\UV{k})\diff\Omega_{\UV{k}}$.

To estimate the NDoF, we assume that the efficiency of the generalized radiation modes separates into efficient modes and inefficient modes. This is similar to the decomposition of the channel eigenvalues $\zeta_n$ in Figs~\ref{fig:idealchannelNDoF} and~\ref{fig:NDoFwaterfilling} which has approximately $\Na$ modes $\zeta_n\Na\approx 1$ and thereafter rapidly decreasing eigenvalues. The normalized modes $\zeta_n$ are independent of the material loss (or similarly sampling density) but the channel eigenvalues increase as the sampling increases.  
The efficiency $\nu_n$ of the generalized radiation modes diminishes rapidly, similar to ordinary radiation modes~\cite{Gustafsson+etal2020}. 
Assuming that the efficient generalized radiation modes have $\nu_n\approx 1$ and inefficient modes have $\nu_n\approx 0$, resulting in efficiencies $\nu_n$ according to
\begin{equation}
 \nu_n \approx 
\begin{cases}
    1 & n < \Na \\
    0 & n > \Na, 
\end{cases}
\label{eq:radmeff}
\end{equation}
where $\Na$ denotes the NDoF for the given shape and frequency. The transition from high to low modal efficiencies is more pronounced for larger electrical sizes, see Figs~\ref{fig:idealchannelNDoF} and~\ref{fig:NDoFwaterfilling}.
Assuming this ideal channel, inserted into~\eqref{eq:sumnu} estimates the NDoF as
\begin{equation}
	\Na =\frac{2A_\T{TR}}{\lambda^2}	
	\quad\text{for } \Na\gg 1,
\label{eq:NDoF_As}
\end{equation}
where the factor of two stems from the two polarizations of the EM field. The corresponding NDoF for 2D cases is
\begin{equation}
	\Na =\frac{L_\T{TR}}{\lambda}	
	\quad\text{for } \Na\gg 1,
\label{eq:NDoF_As2D}
\end{equation}
where $L_\T{RT}$ denotes the mutual shadow length and one polarization degree of freedom is assumed, \cf~\eqref{eq:NDoFshadowArea}.

These generalized radiation modes are slightly different from the modes induced by the decomposition of the channel matrix $\M{H}$ in~\eqref{eq:eigHH}, but are assumed to have similar NDoF. The mutual shadow length $L_\T{TR}$ estimates the NDoF~\eqref{eq:NDoF_As2D} as the shadow length measured in wavelength. This estimate is compared with the numerical evaluation of the channel between $\reg_\T{T}$ and $\reg_\T{R}$. This channel can be modeled in many ways, \eg point sources or expansion of a current density in basis functions. There is also a choice of using electric and/or magnetic current densities (single or double potentials). The presented results are for simplicity compared with the common model based on uniformly sampled ($\varDelta\approx\lambda/5$) point sources~\cite{Piestun+Miller2000}.
To compare different configurations of transmitting and receiving regions, the singular values can be evaluated for a fixed $\Na$ by using~\eqref{eq:NDoF_As2D} to set the wavelengths to
\begin{equation}
    \lambda = L_{\T{TR}}/\Na.
    \label{eq:Na2wavelength}
\end{equation}

\begin{figure}
    \centering
    \includegraphics[width=\figw\columnwidth]{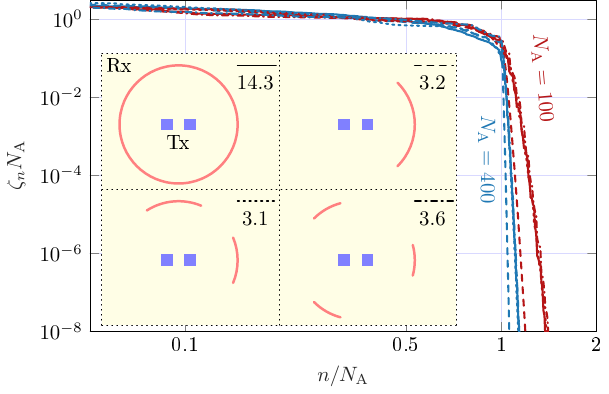}
    \caption{Normalized eigenvalues $\zeta_n$ are shown for a transmitter consisting of two square regions with side length $\ell$, separated by a distance $\ell$. The transmitter is surrounded by a cylindrical receiving region providing full $2\pi$ coverage, as well as partially surrounding regions with $\pi/2$ coverage, as illustrated in the inset. The mode index $n$ is normalized with $\Na \in {100,400}$, and the normalized shadow lengths $L_\T{RT}/\ell$ are indicated in the inset.} 
    \label{fig:NDoFsigma_cyl}
\end{figure}

Normalized singular values $\zeta_n$ for $2\pi$ coverage and three cases of $\pi/2$ coverage are shown in Fig.~\ref{fig:NDoFsigma_cyl}, using wavelengths from~\eqref{eq:Na2wavelength} corresponding to NDoF values $\Na \in {100,400}$. The transmitter is modeled as two square regions with side length $\ell$, separated by a distance $\ell$, with normalized shadow lengths $L_\T{TR}/\ell$ indicated in the inset. 
The shadow length is largest for the surrounding case, $L_\T{TR} \approx 14.3\ell$, whereas the shadow lengths in the other $\pi/2$ cases are approximately one quarter of this value.
All eight cases (four geometries and two NDoF models) exhibit a rapid transition in the eigenvalue spectrum: the eigenvalues remain nearly constant with $\zeta_n\Na \approx 1$ for $n < \Na$, and then decay rapidly for $n > \Na$. This behavior is highlighted by plotting the normalized eigenvalues, $\zeta_n \Na$, versus the normalized index, $n/\Na$.

The presented results are evaluated for receivers in the far field. The same setup is evaluated for a receiver region at a finite distance, \eg a receiver at radius $15\ell$, but the results are indistinguishable. This suggests that the same approach, using rays (plane waves), can be applied to objects at finite distances. 

\section{Communication between volumes and mutual shadow area}\label{S:CommVolMutualShadowArea}
\begin{figure}[t]%
{\centering
\begin{tikzpicture}[scale=1,>= stealth]
	\draw [fill=treg,very thick] plot [smooth cycle,scale=1] coordinates {\Treg};    
	\node at (-0.1,-0.3) {$\reg_\T{T}$};	
	\node at (0.5,0.2) {$\vec{J}_\T{T}(\rv)$};	
\begin{scope}[xshift=5cm]
	\draw [fill=rreg,very thick,rotate=0] plot [smooth cycle,scale=1] coordinates {\Rreg};    
	\node at (0.1,-0.2) {$\vec{E}_\T{R}(\rv)$};	
	\node at (0.2,0.4) {$\reg_\T{R}$};	
\end{scope}	 
\begin{scope}[rotate=10,xshift=5mm,yshift=-2mm]    
\draw[->,thick,decorate,decoration=snake,dgreen] (1.5,-0.5) --  (3.5,-0.4);	
\draw[->,thick,decorate,decoration=snake,dgreen] (1.7,-0.2) --  (3.6,-0.0);	
\draw[->,thick,decorate,decoration=snake,dgreen] (1.6,0.2) --  (3.5,0.3);	
\draw[->,thick,decorate,decoration=snake,dgreen] (1.6,0.6) --  (3.5,0.7);	
\end{scope}

\end{tikzpicture}
\vspace{-1mm}
\par}
\caption{Communication channel represented by waves propagating in a direction $\UV{k}$ between transmitting and receiving regions.}%
\label{fig:TXrays}%
\end{figure}
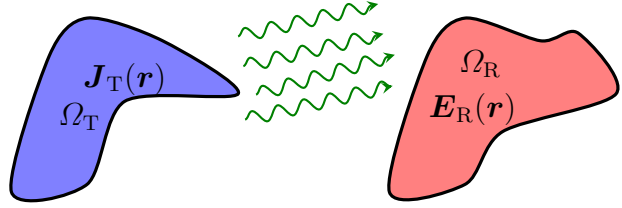

To determine the asymptotic NDoF for communication between two arbitrarily shaped regions, we take a similar approach and focus on the propagation of plane waves between the regions. 
This effectively reduces the question of NDoF to the question about the number of waves that can be supported between the objects. These waves are constructed by a plane-wave expansion of the current density over all directions. This procedure is depicted in Fig.~\ref{fig:TXrays}, where plane waves (rays) in a direction $\UV{k}$ are illustrated. To communicate from $\reg_\T{T}$ to $\reg_\T{R}$ with a ray in the direction $\UV{k}$, the ray needs to intersect both regions. The amount of interaction is proportional to the mutual area of this interaction. This is analogous to the shadow area for the far-field or circumscribing case in~\eqref{eq:NDoFshadowArea} and~\eqref{eq:NDoF_As} which can be interpreted as a total interception of plane waves (rays) from directions described by $\regR$.

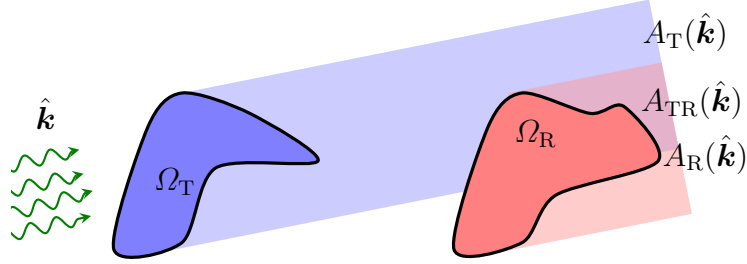
\begin{figure}[t]%
{\centering
\begin{tikzpicture}[scale=0.9,>= stealth]
\def\xs{7.5}
\begin{scope}
    \clip[rotate=11.3] (-0.4,-2.23) rectangle (7.2,1.0);    
    \fill[blue!40!white,opacity=0.5] (0,1) -- (\xs,2.5) -- (\xs,0.22) -- (-0.6,-1.4);  
    \fill[red!40!white,opacity=0.5] (4.8,1) -- (\xs,1.54) -- (\xs,-0.82) -- (4.6,-1.4);	
\end{scope}
    
    \draw [fill=treg,very thick] plot [smooth cycle,scale=1] coordinates {\Treg};    
	\node at (-0.1,-0.3) {$\reg_\T{T}$};	
\begin{scope}[xshift=5cm]
	\draw [fill=rreg,very thick,rotate=0] plot [smooth cycle,scale=1] coordinates {\Rreg};    
	\node at (0.2,0.4) {$\reg_\T{R}$};	
\end{scope}	 
\node at (7.5,0.9) {$\sregTR(\UV{k})$}; 
\node at (7.4,1.9) {$\sregT(\UV{k})$}; 
\node at (7.7,0.1) {$\sregR(\UV{k})$}; 
\begin{scope}[xshift=-45mm,yshift=-10mm]    
\clip (2.0,-0.5) rectangle (4,2.0);        
\begin{scope}[rotate=10,scale=0.9]
\draw[->,thick,decorate,decoration=snake,dgreen] (1.5,-0.5) --  (3.5,-0.4);	
\draw[->,thick,decorate,decoration=snake,dgreen] (1.7,-0.2) --  (3.6,-0.0);	
\draw[->,thick,decorate,decoration=snake,dgreen] (1.6,0.2) --  (3.5,0.3);	
\draw[->,thick,decorate,decoration=snake,dgreen] (1.6,0.6) --  (3.5,0.7);	
\end{scope}
\node at (2.5,1.7) {$\UV{k}$};
\end{scope}
\end{tikzpicture}
\vspace{-1mm}
\par}
\caption{NDoF of the communication channel, represented by the mutual (overlapping) shadow region from transmitting and receiving regions and a plane wave in the direction $\UV{k}$.}%
\label{fig:TXshadow}%
\end{figure}

A simple interpretation of this area comes from the mutual shadow area of two objects, see Fig.~\ref{fig:TXshadow}. In this picture, the transmitting region $\reg_\T{T}$ illuminated by a plane wave in the direction $\UV{k}$ casts a shadow on the receiving region $\reg_\T{R}$. This area can also be seen as the mutual shadow area of the two shadows, \ie the area where the two shadows overlap. This intersection area integrated for all directions is shown here to be proportional to the NDoF between the two regions in the asymptotic limit $\lambda\to 0$.

The total mutual shadow area is determined by integrating the mutual shadow area for all directions $\UV{k}$
\begin{equation}
    L_\T{TR} = \int_0^{2\pi}
    L_\T{TR}(\phi)\diff\phi
    \label{eq:LTR}
\end{equation}
in $\R^2$ and
\begin{equation}
    A_\T{TR} = \int_{4\pi} A_\T{TR}(\UV{k})\diff\Omega_{\UV{k}}
    \label{eq:ATR}
\end{equation}
in $\R^3$, where the shadow region is determined from $\reg_\T{T}$ to $\reg_\T{R}$, \ie the shadow from $\regR$ onto $\regT$ is not included. In~\eqref{eq:LTR} and~\eqref{eq:ATR} $\sregTR(\UV{k})$ denotes the mutual shadow area (length) for a given direction $\UV{k}$ of illumination. 
Here, we also note that the mutual shadow area can alternatively be determined from the difference between the sum of the shadow areas of $\reg_\T{T}$ and $\reg_\T{T}$ minus two times the shadow area of the composed object $\reg_\T{TR}$. The NDoF for communication between systems confined to those regions is approximately
\begin{equation} 
    \Na = 
    \begin{cases}
        L_\T{TR}/\lambda & \text{in }\R^2\\
        A_\T{TR}/\lambda^2 & \text{in }\R^3
    \end{cases}
    \label{eq:NAdef}
\end{equation}
in agreement with~\eqref{eq:NDoF_As2D} and~\eqref{eq:NDoF_As}.
Estimates~\eqref{eq:NAdef} of the asymptotic number of propagating modes for the (scalar) Helmholtz equation are doubled for Maxwell's equations when two orthogonal polarizations contribute. These NDoF reduce to many previously known cases for those specific geometries, \eg the case with a receiving spherical region circumscribing the transmitter~\eqref{eq:NDoFshadowArea} and paraxial approximation~\eqref{eq:NDoFparaxial}.

\begin{figure}
    \centering
    \includegraphics[width=0.5\columnwidth]{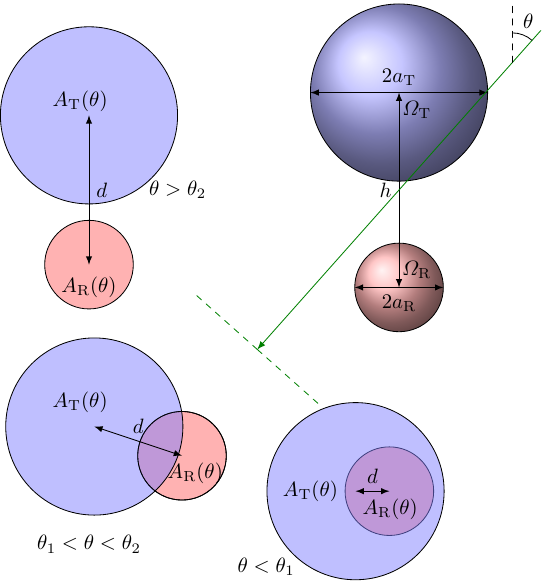}
    \caption{Illustration of the shadow and mutual shadow from two spheres with radii $a_\T{T}$ and $a_\T{R}\leq a_\T{T}$ separated a distance $d$. The spheres are placed on the $\UV{z}$-axis and illuminated from a direction $\UV{k}$ described by the polar angle $\theta$. The shadow of $\reg_\T{R}$ is contained within the shadow of $\reg_\T{T}$ for small $\theta<\theta_1$ with $\sin\theta_1=|a_\T{T}-a_\T{R}|/h$, intersects the shadow for $\theta_1\leq\theta\leq \theta_2$ with $\sin\theta_2=(a_\T{T}+a_\T{R})/h$, and are disjoint for $\theta>\theta_2$, see App.~\ref{S:Sphshadowarea} for details.}
    \label{fig:ShadowAreaSph}
\end{figure}

To illustrate shadow and mutual shadow areas, two spherical regions with radii $a_\T{T} \geq a_\T{R}$ separated by a distance $h$ as depicted in Fig.~\ref{fig:ShadowAreaSph} are considered. The spheres are, without loss of generality, placed on the z-axis, producing a body of revolution (BoR) object with the incident illumination parametrized by the polar angle $\theta$. The shadow areas are determined on a plane perpendicular to the direction of illumination.

The shadow of each sphere is a circular disc with the same radius as the sphere. The two shadows can overlap, partly overlap, or be separated depending on the illumination angle $\theta$ related to the distance $d=h\sin\theta$ between the disc centers. The shadow $\sregR$ is inside the shadow of $\sregT$ for sufficiently small polar angles $\theta<\theta_1$, see App.~\ref{S:Sphshadowarea} for details. This gives a mutual shadow area $\pi a_\T{R}^2$. For larger polar angles $\theta_1<\theta<\theta_2$, the two shadows partly overlap, with a mutual shadow area given in App.~\ref{S:Sphshadowarea}. As the polar angle increases $\theta>\theta_2$, the two shadows separate, and the mutual shadow area vanishes. For non-intersecting shadows, it is not possible to communicate between the regions using rays with this propagation direction.   

As the polar angle $\theta$ approaches $\pi$, there is a corresponding mutual shadow area from the receiving region $\regR$ onto the transmitting region $\regT$. This mutual shadow area has the same size but is not accounted for in the shadow area from $\regT$ to $\regR$ in~\eqref{eq:ATR}. 

The mutual shadow areas $\sregTR(\theta)$ have simple analytical expressions and the total mutual shadow area $A_\T{TR}$ is determined by integration of $\sregTR(\theta)$ over $[0,\theta_2]$ weighted by $2\pi\sin\theta$ from integration over the azimuthal angle $\phi$ of the spherical coordinate system, see App.~\ref{S:Sphshadowarea}. 

The evaluation of shadows is well-developed in computer graphics~\cite{Agoston2005}. Shadows for canonical shapes can be determined semi-analytically, while those for objects described by a mesh can be computed numerically by combining the shadows of individual mesh elements.  

In 3D, shadow areas for arbitrarily shaped regions can be efficiently evaluated using numerical techniques. Consider, for simplicity, a geometry composed of $N$ convex shapes, indexed by $n = 1, \dots, N$, each described by a set of points $\V{r}_{np}$, where $p = 1, \dots, N_\T{p}$. This could, for example, represent a surface mesh with node points. Using spherical coordinates, with the illumination direction defined by the radial vector $\rvh$, we project the shadow onto the orthogonal plane spanned by the unit vectors $\UV{\theta}$ and $\UV{\phi}$, as illustrated in Fig.~\ref{fig:ShadowAreaSph}. 

Standard algorithms, such as the \texttt{convhull} function in MATLAB, can be used to compute the convex hull of these projected points, resulting in polygon-like shapes in 2D. These shapes can then be combined through unions and intersections to facilitate shadow area computations.

The total mutual shadow area, weighted by the number of boundary-intersecting rays, is also given by the surface integral  
\begin{equation}
    A_\T{TR}=
    \int_{\regT}\int_{\regR} \frac{ |\UV{n}_\T{T} \cdot \V{R}|\ |\UV{n}_\T{R} \cdot \V{R}|}{|\V{R}|^4 \ \xi_\T{T}(\V{R})\ \xi_\T{R}(\V{R})}\diffS_\T{R} \diffS_\T{T},
    \label{eq:MutualShadowArea}
\end{equation}  
where $\V{R} = \V{r}_\T{R} - \V{r}_\T{T}$ represents the vector connecting points on the receiver and transmitter surfaces, $\UV{n}_\T{X}$ denotes the unit normal, and $\xi_\T{X}(\V{R})$ represents the number of surface crossings of $\reg_\T{X}$ for a ray in the direction $\V{R}$. For convex volumetric regions, $\xi_\T{X} = 2$, while for planar regions, \eg flat surfaces, $\xi_\T{X} = 1$. This integral sums the contributions of infinitesimal elements under the paraxial approximation~\eqref{eq:NDoFparaxial} and can be efficiently evaluated using a numerical mesh.  

\begin{figure}
\centering
\begin{tikzpicture}
    \def\ax{3}
    \def\ay{5}
    \def\bx{4}
    \def\by{3}
    \def\cx{1}
    \def\cy{2}
    \def\dx{2}
    \def\dy{4}
    \def\pp{4pt}
    \begin{scope}[rotate=200]        
    \fill[blue!40!white,opacity=0.5] (\ax,\ay) -- (\bx,\by) -- (0,\by) -- (0,\ay);    
    \fill[red!40!white,opacity=0.5] (\cx,\cy) -- (\dx,\dy) -- (0,\dy) -- (0,\cy);'
    
    \draw[very thick,dblue] (\ax,\ay) node[circle,fill=dblue,inner sep=0pt,minimum size=\pp] {} -- (\bx,\by) node[circle,fill=dblue,inner sep=0pt,minimum size=\pp] {};
    \node[below] at (\ax,\ay) {$\V{r}_{1,1}$};
    \node[above] at (\bx,\by) {$\V{r}_{1,2}$};
    \draw[dashed,dblue] (0,\ay) node[circle,fill=dblue,inner sep=0pt,minimum size=\pp] {} -- (0,\by) node[circle,fill=dblue,inner sep=0pt,minimum size=\pp] {};  
    \draw[very thick,dred] (\cx,\cy) node[circle,fill=dred,inner sep=0pt,minimum size=\pp] {} -- (\dx,\dy) node[circle,fill=dred,inner sep=0pt,minimum size=\pp] {};    
    \node[above] at (\cx,\cy) {$\V{r}_{2,1}$};
    \node[below] at (\dx,\dy) {$\V{r}_{2,2}$};
    \draw[dashed,dred] (0,\cy) node[circle,fill=dred,inner sep=0pt,minimum size=\pp] {} -- (0,\dy) node[circle,fill=dred,inner sep=0pt,minimum size=\pp] {};
    \node[right] at (0,\ay) {$p_{1,1}$};
    \node[right] at (0,\by) {$p_{1,2}$};
    \node[right] at (0,\cy) {$p_{2,1}$};
    \node[right] at (0,\dy) {$p_{2,2}$};
    \draw[->] (0,6.1) -- node[right] {$\UV{p}$} +(0,-0.7);
    
    \node[dblue] at (4,4) {$\reg_\T{T}$};
    \node[dred] at (1.5,2) {$\reg_\T{R}$};
    \node[dblue] at (-0.6,4.6) {$\sregT(\UV{k})$};
    \node[dred] at (-0.6,2.6) {$\sregR(\UV{k})$};
    \node at (-0.8,3.7) {$\sregTR(\UV{k})$};
    
\begin{scope}[xshift=80mm,yshift=32mm] 
\begin{scope}[rotate=180]    
\draw[->,thick,decorate,decoration=snake,dgreen] (1.5,-0.5) --  +(1.5,0);	
\draw[->,thick,decorate,decoration=snake,dgreen] (1.5,0) --  +(1.5,0);	
\draw[->,thick,decorate,decoration=snake,dgreen] (1.5,0.5) --  +(1.5,0);	
\end{scope}
\end{scope}    
\end{scope}

\begin{scope}[xshift=-4.5cm,yshift=-4cm]
\draw[->] (0,0) -- (1,0) node[below] {$\UV{x}$};
\draw[->] (0,0) -- (0,1) node[left] {$\UV{y}$};
\draw[->] (0,0) -- (0.8,0.5) node[above] {$\UV{k}$};    
\end{scope}

\end{tikzpicture}

\caption{Multual shadow length between two line objects in $\R^2$ illuminated from direction $\UV{k}$ described by an angle $\phi$. The shadow is projected on a line perpendicular to $\UV{k}$.}
\label{fig:LinesShadowGeo}
\end{figure}
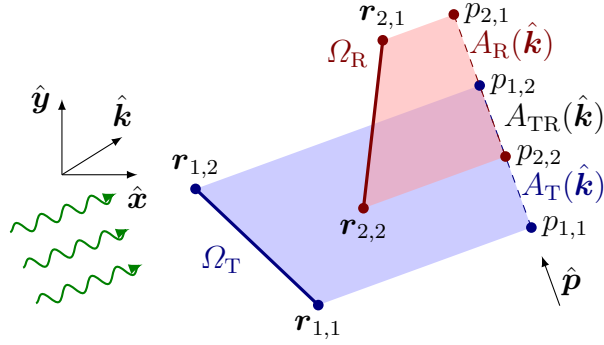

In 2D, shadow lengths are straightforward to evaluate. Consider two line objects, as depicted in Fig.~\ref{fig:LinesShadowGeo} and detailed in App.~\ref{S:Geo2Dlines}. The shadow of a line object forms an interval whose endpoints are determined by projecting the object’s endpoints onto a line $\UV{p}$ perpendicular to the illumination direction, \ie $\UV{p} \cdot \UV{k} = 0$. The intersection length is then determined from the overlap of these shadow intervals (see App.~\ref{S:Geo2Dlines} for details).  

The total mutual shadow length, weighted similarly to~\eqref{eq:MutualShadowArea} by the number of boundary-intersecting rays, can be evaluated as a line integral
\begin{equation}
    L_\T{TR}=
    \int_{\regT}\int_{\regR} \frac{ |\UV{n}_\T{T}\cdot\V{R}|\ |\UV{n}_\T{R}\cdot\V{R}|}{|\V{R}|^3 \ \xi_\T{T}(\V{R})\ \xi_\T{R}(\V{R})}\diff l_\T{R}\diff l_\T{T}.
    \label{eq:MutualShadowLength}
\end{equation}

\section{NDoF and shadow length in 2D}\label{S:2D}
\begin{figure}
    \centering
    \includegraphics[width=\figw\columnwidth]{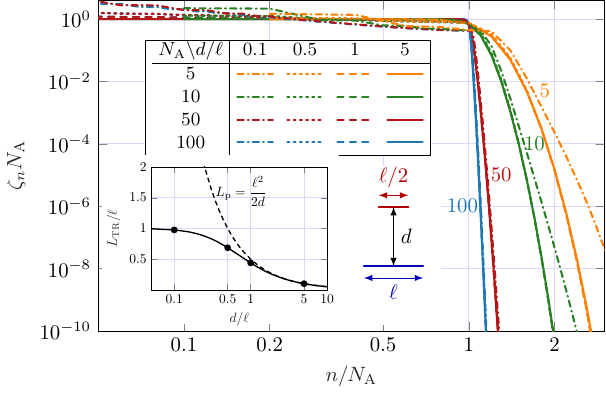}
    \caption{Normalized singular values $\zeta_n$ for two lines with length $\ell$ and $\ell/2$ separated a distance $d$. The mode index $n$ is normalized with $\Na$ determined from the mutual shadow length~\eqref{eq:SL_twolines}.}
    \label{fig:NDoFsigma2lines}
\end{figure}

To illustrate NDoF and shadow length, consider first two parallel lines with lengths $\ell_1$ and $\ell_2$ separated by a distance $d$ as shown in the inset of Fig.~\ref{fig:NDoFsigma2lines}. The mutual shadow length is determined by integrating $\sregTR(\phi)$ over all illumination directions $\phi$, with the transmitter casting a shadow on the receiver  
\begin{equation}
    2\int_0^{\phi_1}\ell_2\cos\phi\diff\phi
    +2\int_{\phi_1}^{\phi_2}\frac{\ell_1+\ell_2}{2}\cos\phi-d\sin\phi\diff\phi
\end{equation}
with $\tan\phi_1=\delta$ and $\tan\phi_2=\beta$, where $\delta=|\ell_1-\ell_2|/(2d)$ and $\beta=(\ell_1+\ell_2)/(2d)$. Integration over $\phi$ leads to the total shadow length
\begin{equation}
    L_{\T{TR}}=2d\big(\sqrt{1+\beta^2} - \sqrt{1+\delta^2}\big)
    \approx 
    \begin{cases}
    2\min\{\ell_1,\ell_2\},\ d\to 0 \\
    \ell_1\ell_2/d,\  d\to\infty, 
    \end{cases}
    \label{eq:SL_twolines}
\end{equation}
where the paraxial result~\eqref{eq:NDoFparaxial} is recovered as $d\to\infty$ and Weyl's law for $d\ll \ell$, with a factor $1/2$ due to radiation in one direction. This expression is also consistent with the square root of the results in~\cite[Eq. (32)]{Maisto+etal2021}.

Different configurations of transmitting and receiving regions are compared by evaluating the eigenvalues of the channel correlation matrix for a fixed $\Na$ by setting the wavelength to $\lambda=L_{\T{TR}}/\Na$ and using uniformly sampled point sources with $\varDelta\approx\lambda/5$~\cite{Piestun+Miller2000}.

The results in Fig.~\ref{fig:NDoFsigma2lines} show the normalized eigenvalues $\zeta_n$ in~\eqref{eq:HHeigNorm} of the resulting channel matrix for $\Na\in\{5,10,50,100\}$ and distances $d/\ell\in\{0.1,0.5,1,5\}$. 
The NDoF $\Na$ is evaluated using~\eqref{eq:SL_twolines} as depicted in the inset of Fig.~\ref{fig:NDoFsigma2lines}, where the considered cases $d/\ell\in\{0.1,0.5,1,5\}$ are indicated by markers. Small distances have mutual shadow lengths $L_\T{TR}\approx \ell$ corresponding to twice the shorter strip length~\eqref{eq:SL_twolines}. The shadow length decreases with the distances and approximates the paraxial approximation as $d/\ell\to\infty$. For $d/\ell = 5$, the shadow length $L_\T{TR}\approx \ell/10$ implies a tenfold decrease in the NDoF for a fixed wavelength compared to the $d\ll\ell$ case. The electrical sizes varies between $\ell/\lambda\approx 5$ for $\Na=5$ and $d/\ell=0.1$ to $\ell/\lambda\approx 1000$ for $\Na=100$ and $d/\ell=5$. The singular value index $n$ on the horizontal axis is scaled with the NDoF $\Na$. 

The normalized eigenvalues $\zeta_n$ are approximately equal to $1/\Na$ for $n<\Na$. A 'corner' in the singular values is observed around $n\approx \Na$. For indices $n>\Na$, the singular values $\zeta_n$ decay rapidly, particularly for large $\Na$. 
This sharp transition from an almost constant $\zeta_n\Na\approx 1$ to a rapidly decaying $\zeta_n$ is used to define a NDoF, here denoted $\Nk$. This implies that there are $\Nk$ modes with approximately the same performance, and the performance decreases rapidly if additional modes are used, indicating a very high cost to use more than $\Nk$ modes. 
The results show that this 'corner' becomes more pronounced as the electrical size increases, or more precisely, as the NDoF increases.

The results in Fig.~\ref{fig:NDoFsigma2lines} indicate a strong correlation between the NDoF estimates from the shadow area $\Na$ and the corner of the eigenvalue distribution $\Nk$. They also show that the distribution of the eigenvalues $\zeta_n$ can be approximated with an ideal distribution consisting of $\Na$ modes with amplitude $1/\Na$. The curves for the same $\Na$ but different distances $d/\ell$ overlap, except for the short distance $d=0.1\ell$ and low $\Na\in\{5,10\}$. These cases correspond to a separation between the transmitting and receiving regions  $d/\lambda\approx\Na d/\ell\in\{0.5,1\}$, which is on the order of a wavelength. The results coincide for larger $\Na$ corresponding to shorter wavelengths, supporting $\Na$ as the NDoF as $\lambda\to 0$.  For sub-wavelength distances, the NDoF can be higher due to strong reactive coupling~\cite{Ji+etal2023}.

\begin{figure}
    \centering
    \includegraphics[width=\figw\columnwidth]{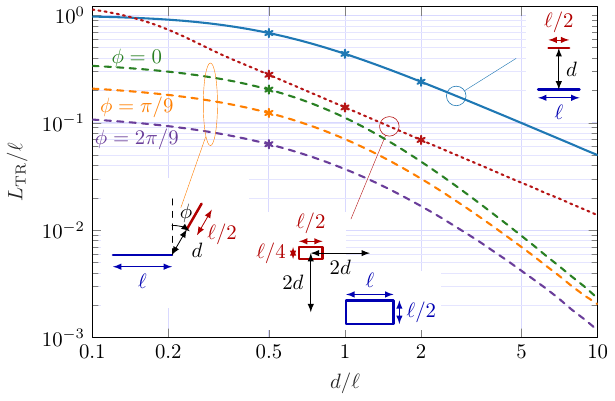}
    \caption{Normalized mutual shadow lengths $L_\T{TR}/\ell$ for parallel lines (solid), rotated lines (dashed), and separated rectangles (dotted).}
    \label{fig:NDoF2D_geos}
\end{figure}

The broadside configuration maximizes the mutual shadow length. Rotating one of the lines reduces this length. Consider a setup with two lines, one of length $\ell$ and the other $\ell/2$, separated by a distance $\ell/2$ and rotated by an angle $\phi$, as shown in the inset of Fig.~\ref{fig:NDoF2D_geos}. The mutual shadow length is evaluated as described in App.~\ref{S:Geo2Dlines}. The mutual shadow length vanishes for the endfire configuration ($\phi = \pi/2$), at which the estimate~\eqref{eq:NAdef} becomes invalid and numerical evaluations indicate a single DoF. The mutual shadow lengths $L_\T{TR}$ are plotted in Fig.~\ref{fig:NDoF2D_geos} for $\phi\in\{0,\pi/9,2\pi/9\}$. Additionally, a setup with rectangular regions is included in Fig.~\ref{fig:NDoF2D_geos}.

\begin{figure}
    \centering
    \includegraphics[width=\figw\columnwidth]{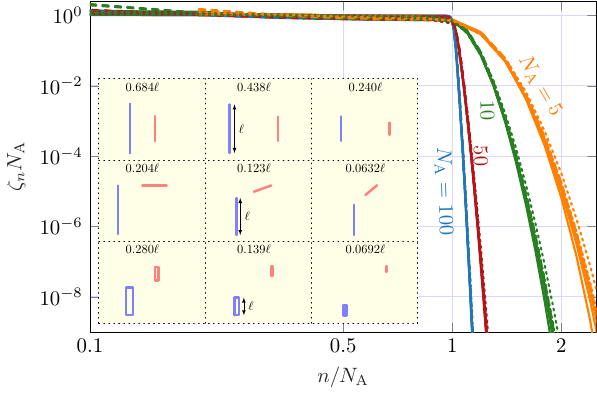}
    \caption{Normalized eigenvalues $\zeta_n$ for the nine different configurations shown in the inset using wavelengths $\lambda=L_\T{TR}/\Na$. The mode index is normalized with $\Na$. Parallel lines as in Fig.~\ref{fig:NDoFsigma2lines} (solid curves), rotated lines (dashed lines), and rectangles (dotted lines) with separation distances indicated by the line thickness, see markers in Fig.~\ref{fig:NDoF2D_geos}.}
    \label{fig:NDoFsigma_norm}
\end{figure}

Normalized singular values $\zeta_n$ are depicted in Fig.~\ref{fig:NDoFsigma_norm} for $\Na\in\{5,10,50,100\}$ and the nine configurations shown in the inset and in Fig.~\ref{fig:NDoF2D_geos}. Normalizing by $\Na$ groups the curves for the different configurations, indicating that the mutual shadow length is a fundamental parameter for the NDoF and the distribution of the eigenvalues. The spread within each group decreases as $\Na$ increases, and the different configurations for the $\Na=100$ case overlap for $n>\Na$. The special case with two perpendicular lines is also analyzed in~\cite{Pierri+Moretta2021}.

The results in Fig.~\ref{fig:NDoFsigma_norm} depend slightly on the specific configuration and the point source model used for the channel matrix. This dependence can be attributed to the symmetry of the point source model, where the radiated field from a point source is isotropic. Consequently, the radiated field from a line region is symmetric in directions away from the line, reducing the NDoF. However, this symmetry is broken for regions with an inner region, such as an area in 2D or a volume in 3D. The symmetry is also broken by using both electric and magnetic current sources for electromagnetic fields~\cite{Gustafsson2025a}, similar to single and double-layer potentials for scalar fields.

\section{NDoF and shadow area in 3D}\label{S:3D}

The mutual shadow area for two parallel discs with radius $a$ separated by a distance $d$ can be evaluated analytically using~\eqref{eq:MutualShadowArea} together with the fact each ray intersects each region once ($\xi_\T{X}=1$) giving the total mutual shadow area
\begin{equation}
    A_\T{RT} = \frac{\pi^2}{4} \big(\sqrt{4a^2+d^2}-d\big)^2
    \approx\begin{cases}
        \pi^2 a^2=\pi A & d\to 0\\
        \pi^2 a^4/d^2=A^2/d^2 & d\to\infty,
    \end{cases}
    \label{eq:MutualSA2Discs}
\end{equation}
where $A=\pi a^2$ denotes the disc area. Weyl's law is retrieved as $d\to 0$ and the paraxial approximation~\eqref{eq:NDoFparaxial} as $d\to\infty$.

\begin{figure}
    \centering
    \includegraphics[width=\figw\columnwidth]{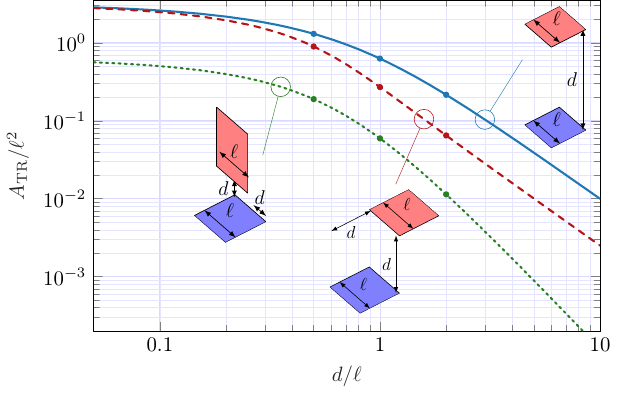}
    \caption{Mutual shadow area in units of $\ell^2$ for communication between two square regions with side length $\ell$ using setups as depicted in the insets.}
    \label{fig:ShadowAreaR2R}
\end{figure}

Mutual shadow area for three configurations using square plates with side lengths $\ell$ are depicted in Fig.~\ref{fig:ShadowAreaR2R}. Parallel plates separated a distance $d$, similar to the setup in Fig.~\ref{fig:idealchannelNDoF}, are shown with a solid line. The mutual shadow area approaches $A_\T{TR}\approx \pi\ell^2$ for $d\ll \ell$ according to half of Weyl's law. For larger distances, it approaches the paraxial approximation~\eqref{eq:NDoFparaxial} $A_\T{TR}\approx \ell^4/d^2$. A similar set-up is analyzed in~\cite[Eq. (32)]{Maisto+etal2021}, but there assuming the product of the 2D results~\eqref{eq:SL_twolines} producing slightly different results.

Combining the vertical separation with a transverse shift is illustrated by the dashed curve. The effect of the transverse shift is negligible compared with the vertical shift for $d\ll\ell$, but its impact increases for $d> \ell$. This is partly explained by the reduced shadow due to the projection of the planar regions along a $45^{\circ}$ angle and the increased distance, which together reduce the mutual shadow area approximately a factor of 4 for $d\gg \ell$. The dotted line illustrates the case with one rectangle rotated $90^{\circ}$, demonstrating an end-fire configuration. This reduces the mutual shadow area, both for small and large distances. Note that the mutual shadow area vanishes for an end-fire setup of two plates in the same plane, producing a scaling similar to the 2D case.

\begin{figure}
    \centering
    \includegraphics[width=\figw\columnwidth]{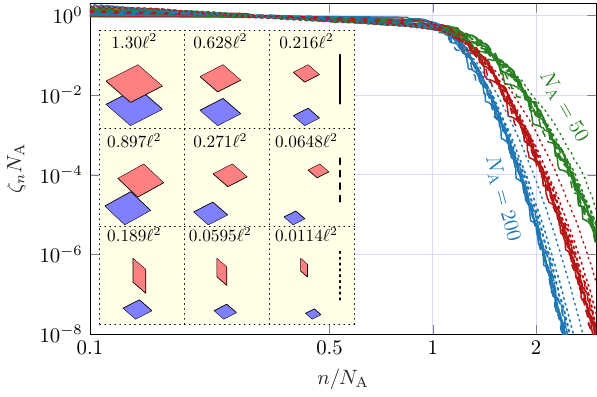}
    \caption{Normalized eigenvalue values $\zeta_n$ of the channel matrix for communication between nine configurations for two square regions with side length $\ell$ evaluated for $\Na\in\{50,100,200\}$. The setups are depicted in the inset together with their mutual shadow area in units of $\ell^2$ for the distances $d/\ell\in\{0.5,1,2\}$ in Fig.~\ref{fig:ShadowAreaR2R}.}
    \label{fig:NDoFsvdR2R}
\end{figure}

Normalized eigenvalues $\zeta_n$ for the setups in Fig.~\ref{fig:ShadowAreaR2R} using $d/\ell\in\{0.5,1,2\}$ and $\Na\in\{50,100,200\}$ are shown in Fig.~\ref{fig:NDoFsvdR2R}. The different setups are separated by solid, dashed, and dotted line styles as in Fig.~\ref{fig:ShadowAreaR2R} and line widths for distances. The results show that $\zeta_n\Na\approx 1$ for $n<\Na$ and that the curves cluster for different $\Na$ for $n>\Na$. The slopes of the eigenvalues are less steep than for the corresponding 2D cases in Fig.~\ref{fig:NDoFsigma_norm}. This is partly explained by the distribution of the DoF over two orthogonal directions. 

Evaluation of the channel for fixed NDoF $\Na$ case corresponds to evaluation of the channel for different wavelengths. The top left case in Fig.~\ref{fig:NDoFsvdR2R} has a mutual shadow area of $A_\T{TR}\approx 1.3\ell^2$ which for the $\Na=200$ case implies $\ell^2/\lambda^2=\Na\ell/A_\T{TR}\approx 154$. Using 5 sample points per wavelength produces a channel matrix with dimension $N_\T{T}=N_\T{R}\approx 4\ 10^3$. This is easily handled on a standard PC. The corresponding bottom right case has a much smaller mutual shadow area $A_\T{TR}\approx 0.0114\ell^2$ which produces a channel matrix of dimension $N_\T{T}=N_\T{R}\approx 4\ 10^5$. 

For these large dimensions, storage of $\M{H}$ and evaluation of singular values or eigenvalues are challenging. However, using the low-rank property of the channel matrix $\M{H}$, \ie the $\T{rank}(\M{H})\approx \Na\ll \min\{N_\T{T},N_\T{R}\}$ is much smaller than the dimension of $\M{H}$ opens for, \eg evaluations based on randomized SVD~\cite{Halko+etal2011}.

Randomized SVD is an effective approach to evaluate the dominant singular values of low-rank matrices~\cite{Halko+etal2011} and can be applied to evaluate $\svd(\M{H})$, and hence $\zeta_n$. Moreover, the randomized SVD is matrix-free, meaning that the full channel matrix is not needed for the evaluation of the singular values. This makes it suitable for cases with very large dimensions of $\M{H}$, when $\lambda \to 0$. The matrix-free randomized SVD can be interpreted as evaluating the received signals for a set of random excitations.

The rank of the channel matrix can be approximated by $\Na$. By constructing a $N_\T{T}\times P$ random matrix $\M{A}$ with, for example, $P = 3\Na$, the received signals $\M{Y} = \M{H}\M{A}$ can be evaluated, where $\M{Y}$ has dimensions $N_\T{T}\times P$, without evaluation of $\M{H}$. An orthonormal basis $\M{W}$ of $\M{Y}$ is then retransmitted through the adjoint channel, which is evaluated by exciting the channel from the receiver side using $\M{W}^{\ast}$, \ie $\M{B} = \M{W}^\herm \M{H} = (\M{H}^{\trans} \M{W}^{\ast})^{\trans}$. This final matrix $\M{B}$ is of size $P \times P$ and is therefore easy to evaluate using an ordinary SVD, \ie $\svd(\M{B})$. Its singular values approximate the dominant singular values of $\svd(\M{H})$~\cite{Halko+etal2011} and hence square roots the eigenvalues $\sigma_n$ in~\eqref{eq:HHeigNorm}.

\section{Conclusions}\label{S:Conclusions}
This study demonstrates that the asymptotic NDoF for communication between regions is approximated by the mutual shadow area of the regions when measured in wavelengths. This analytical estimate provides valuable physical insights and complements numerical evaluations. The numerical results reveal that the eigenvalues of the channel are approximately constant up to the NDoF, after which they decay rapidly for the presented cases. Additionally, it is observed that the normalized eigenvalues of the channel correlation matrix tend to cluster for different setups with the same NDoF. The use of randomized singular value decomposition enables the evaluation of eigenvalues for the correlation matrix, facilitating simulations for setups with small wavelengths. The presented theory and results enhance the understanding of communication system performance in various setups of transmitter and receiver geometries in a free-space propagation environment.

The results presented here primarily focus on simple setups involving lines and plates. However, the approach is versatile and can be applied to arbitrarily shaped regions, including non-convex and non-connected geometries. In setups with symmetries, it may be necessary to consider both electric and magnetic current densities. The generalized radiation modes, derived from the radiated field over a finite far-field region, are also directly applicable to the study of characteristic modes. This framework facilitates the examination of the NDoF for fixed structures, offering valuable insights into their performance and behavior.

\appendices
\section{Electromagnetic modeling}\label{S:currentdensity}
The current density $\V{J}(\rv)$ in $\regT$ is expanded in a set of basis functions $\V{\psi}_n(\rv)$~\cite{Harrington1968}
\begin{equation}
    \V{J}(\rv) = \sum_{n=1}^{N_\T{T}} I_n\V{\psi}_n(\rv),
    \label{eq:Jbasis}
\end{equation}
with the scalar expansion coefficients $I_n$ collected in a column matrix $\M{I}$.
These basis functions can be local, such as divergence conforming or point sources, or global, such as spherical waves. The least-squares, $\T{L}^2$, norm of the current density is
\begin{equation}
    \int |\V{J}|^2\diffV
    =\sum_{mn} I_m^{\ast}I_n\int \V{\psi}_m(\rv)\cdot\V{\psi}_n(\rv)\diffV
    =\M{I}^{\herm}\V{\Psi}\M{I},
\label{eq:Jnorm}
\end{equation}
where $\V{\Psi}$ denotes the Gram matrix~\cite{Horn+Johnson1991}. The $\T{L}^2$ norm is not defined for point sources, but for sub-wavelength sampling, it is possible to consider the point sources as an approximation of pulse basis functions, \ie constant basis functions with unit volume (area). For the case with uniform sampling using point sources, the Gram matrix is proportional to the identity matrix. Homogeneous material losses modeled with a resistivity $\rho_\T{r}$ are described by a material loss matrix $\M{R}_\rho=\rho_\T{r}\V{\Psi}$, meaning that the current norm~\eqref{eq:Jnorm} and material losses are related.   

The radiated field from the current density is determined by a convolution with a Green's function. 
The used scalar Green's function in $\R^2$ and $\R^3$ are
\begin{equation}
    G_2(\V{r}-\V{r}') = \frac{\ju}{4}\T{H}^{(2)}_0(k|\V{r}-\V{r}'|)
    \quad\text{and }
    G_3 = \frac{\eu^{-\ju k|\V{r}-\V{r}'|}}{4\pi|\V{r}-\V{r}'|},
\end{equation}
respectively, where $\T{H}^{(2)}_0$ denotes a Hankel function and the Green's dyadic is $\M{G}=(\Id+k^{-2}\nabla\nabla)G_3$.

\section{Capacity and generalized radiation modes}\label{S:CapRadmodes}
The capacity of the channel from $\regT$ to $\regR$ is determined by expanding the currents in $\regT$ in basis functions~\eqref{eq:Jbasis} and sampling the field over $\regR$. 
The field is power normalized and involves a quadrature scheme with quadrature weights $\Lambda_p^2$ for quadrature points (directions) $\UV{k}_p$ and polarizations $\UV{e}_p$ with $p\in[1,N_\T{R}]$. We use a spherical (or cylindrical) coordinate system with directions $\UV{k}_p$ described by polar $\theta_p$ and azimuthal $\phi_p$ coordinates and polarizations in the $\UV{\theta}_p$ and $\UV{\phi}_p$ directions~\cite{Kristensson2016}. 
The quadrature weights can also be interpreted as an SNR weight or spectral density for a communication system, which also includes point measures for receivers in discrete directions. Moreover, it is also assumed that sufficiently many quadrature points are used such that the numerical quadrature error is negligible.

Collect the received field in a column matrix $\M{f}=\M{\Lambda}\M{F}\M{I}+\M{n}$, where $\M{F}$ is a matrix describing the linear map from the currents to the received field in the quadrature points $\UV{k}_p$ and polarizations $\UV{e}_p$~\cite{Capek+etal2023a} including the square root of the quadrature weights collected in the diagonal matrix $\M{\Lambda}$ and additive noise $\M{n}$. The capacity (spectral efficiency) of this communication system $\M{f}=\M{\Lambda}\M{F}\M{I}+\M{n}$, \ie the channel $\M{H}=\M{\Lambda}\M{F}$, constrained by $\Tr(\M{R}_\T{x}\M{P})=1$ is determined by the optimization problem~\cite{Ehrenborg+Gustafsson2020,Gustafsson2025a}
\begin{equation}
\begin{aligned}
& \maximize &&  \log_2\big(\det(\Id+\gamma\M{\Lambda}\M{F}\M{P}\M{F}^{\herm}\M{\Lambda})\big) \\
& \subto && \Tr(\M{R}_\T{x}\M{P})=1 \\
& && \M{P} \succeq \M{0},
\end{aligned}
\label{eq:MaxCapI1}
\end{equation}
where $\M{P}=\mathcal{E}\{\M{I}\M{I}^{\herm}\}$ denotes the covariance matrix of the currents,  $\gamma$ quantifies the SNR from the additive noise $\M{n}$, $\Id$ denotes the identity matrix, and $\succeq$ positive semi definite. In~\eqref{eq:MaxCapI1}, the current is without loss of generality normalized to a fixed value. The power constraint $\Tr(\M{R}_\T{x}\M{P})=1$ with $\M{R}_\T{x}$ a symmetric positive definite matrix is used to model different communication situations. Considering, \eg the dissipated power $\frac{1}{2}\M{I}^{\herm}\M{R}\M{I}$ in radiation and material losses uses the method of moments (MoM) resistance matrix~\cite{Harrington1968} $\M{R}_\T{x}=\M{R}$.
The resistance matrix $\M{R}$ can be decomposed into its radiation $\M{R}_0$ and material part $\M{R}_\rho$~\cite{Harrington1968,Gustafsson+etal2020}, where $\frac{1}{2}\M{I}^{\herm}\M{R}_0\M{I}$ and $\frac{1}{2}\M{I}^{\herm}\M{R}_\rho\M{I}$ model radiated power and power dissipated in ohmic or dielectric losses. Constraining only material losses uses $\M{R}_\T{x}=\M{R}_\rho$  in~\eqref{eq:MaxCapI1}. These material losses can also be interpreted as a constraint on the amplitude of the current density for a region with homogeneous losses modelled by a resistivity $\rho_\T{r}$ and the loss matrix  $\M{R}_\rho=\rho_\T{r}\V{\Psi}$ with the Gram matrix~\cite{Horn+Johnson1991} $\V{\Psi}$ and least-squared norm of the current density $\M{I}^{\herm}\V{\Psi}\M{I}$, see App.~\ref{S:currentdensity}. Alternative formulations of~\eqref{eq:MaxCapI1} include constraints on efficiency~\cite{Ehrenborg+Gustafsson2018,Ehrenborg+Gustafsson2020} or bandwidth from the reactive energy around the antenna~\cite{Ehrenborg+Gustafsson2018,Ehrenborg+etal2021}. 

The capacity~\eqref{eq:MaxCapI1} is determined by the water-filling algorithm~\cite{Paulraj+etal2003} for which it is convenient to rewrite~\eqref{eq:MaxCapI1} by a change of variable $\tM{P}=\M{G}\M{P}\M{G}^{\herm}$, where $\M{R}_\T{x}=\M{G}^{\herm}\M{G}$ is a factorization of a symmetric positive definite matrix~\cite{Horn+Johnson1991}. Substituting $\tM{P}$ into~\eqref{eq:MaxCapI1} results in
\begin{equation}
\begin{aligned}
& \maximize &&  \log_2\big(\det(\Id+\gamma\M{H}\tM{P}\M{H}^{\herm})\big) &&&\\
& \subto && \Tr(\tM{P})=1 &&&\\
& && \tM{P} \succeq \M{0}&&&
\end{aligned}
\label{eq:MaxCapI2}
\end{equation}
with the channel matrix $\M{H}=\M{\Lambda}\M{F}\M{G}^{-1}$. This optimization problem is diagonalized by an SVD of the channel matrix $\M{H}$, \ie square roots of the eigenvalues of $\M{H}^{\herm}\M{H} = \M{G}^{-\herm}\M{F}^{\herm}\M{\Lambda}^2\M{F}\M{G}^{-1}$, which can be written $\M{G}^{-\herm}\M{F}^{\herm}\M{\Lambda}^2\M{F}\M{G}^{-1}\M{V}_n = \nu_n\M{V}_n$. Simplifying by multiplication with $\M{G}^{\herm}$ and setting $\M{I}_n=\M{G}^{-1}\M{V}_n$ results in the eigenvalue problem
\begin{equation}  
\M{F}^{\herm} \M{\Lambda}^2 \M{F} \M{I}_n = \nu_n \M{R}_\T{x} \M{I}_n.  
\label{eq:RadmEff}  
\end{equation}  
These modes generalize the radiation modes~\cite{Schab2016,Ehrenborg+Gustafsson2020,Gustafsson2025a} from a receiving structure surrounding the transmitter to an arbitrary receiver region. For simplicity, we refer to the modes $\M{I}_n$ in~\eqref{eq:RadmEff} as generalized radiation modes.  
These modal currents~\eqref{eq:RadmEff} are orthogonal with respect to the resistance matrix $\M{R}_\T{x}$ and the region $\regR$
\begin{equation}
 \M{I}_m^{\herm}\M{R}_\T{x}\M{I}_n=\delta_{mn}
 \quad\text{and }
\M{I}_m^{\herm}\M{F}^{\herm}\M{\Lambda}^2\M{F}\M{I}_n=\nu_n\delta_{mn}.
\label{eq:radmOrthogonal}
\end{equation}
Generalized radiation modes~\eqref{eq:RadmEff} maximize the (Rayleigh) quotient between the radiated power over the directions $\regR$ and dissipated power quantified by $\M{R}_\T{x}$ 
\begin{equation}
	\frac{\mrm{radiated\ power\ in\ } \regR}{\mrm{dissipated\ power}}
	=
	\frac{\M{I}_m^{\herm}\M{F}^{\herm}\M{\Lambda}^2\M{F}\M{I}_n}{\M{I}_m^{\herm}\M{R}_\T{x}\M{I}_n}
	=\delta_{mn}\nu_n.
\label{eq:Rayleighradmod}
\end{equation}
Using $\M{R}_\T{x}=\M{R}$ interprets the generalized radiation mode eigenvalues $\nu_n\in[0,1]$ as an efficiency for orthogonal currents in $\regT$ where only radiation in the region $\regR$ is accounted for. 

\section{Trace and power identities}\label{S:trace}
The analytical solution is determined by relating the sum of the efficiencies $\nu_n$ to geometrical properties of $\regT$ and $\regR$ using the maximal effective area of antennas designed within $\regT$. The sum of the efficiencies $\nu_n$ is given by the trace~\cite{Horn+Johnson1991} 
\begin{equation}    
    \sum_{n=1}^N \nu_n=\Tr(\M{G}^{-\herm}\M{F}^{\herm}\M{\Lambda}^2\M{F}\M{G}^{-1})
    =\Tr(\M{R}_\T{x}^{-1}\M{F}^{\herm}\M{\Lambda}^2\M{F}) 
    =\Tr(\M{F}\M{R}_\T{x}^{-1}\M{F}^{\herm}\M{\Lambda}^2)
    =\sum_{p=1}^{N_\T{R}}\Lambda_p^2\M{F}_{p}\M{R}_\T{x}^{-1}\M{F}_{p}^{\herm},
    \label{eq:sumefficiency}
\end{equation}
where cyclic permutations of the trace operator $\Tr$ are used and $\M{F}_{p}$ denotes row $p$ of $\M{F}$. 
The identity~\eqref{eq:sumefficiency} is valid for arbitrary constraints $\M{R}_\T{x}$ in~\eqref{eq:MaxCapI1}. Here, we consider $\M{R}_\T{x}=\M{R}$ to constrain the dissipated power (radiation and losses) and $\M{R}_\T{x}=\M{R}_\rho$ to constrain material losses (or equivalently the norm of the current density). 

The right-hand side of~\eqref{eq:sumefficiency} is related to the partial effective area $A_{\mrm{eff}}=\lambda^2 G/(4\pi)$, with the partial gain $G$ in a direction $\UV{k}$ and polarization $\UV{e}$ is given by the partial radiation intensity normalized by the dissipated power and $4\pi$~\cite{Harrington1968,IEEEantennaterms2013}. Expressing the gain and effective area in the current matrix $\M{I}$, resistance matrix $\M{R}$, and radiation matrix $\M{F}_p$ yields a framework useful for optimization over the currents. The maximal partial effective area for antennas fitting within $\regT$ is determined from the solution of the optimization problem~\cite{Gustafsson+Capek2019}
\begin{equation}
\begin{aligned}
& \maximize && A_{\mrm{eff},p}=\lambda^2\M{I}^{\herm}\M{F}_p^{\herm}\M{F}_p\M{I}  \\
& \subto && \M{I}^{\herm}\M{R}\M{I}=1,
\end{aligned}
\label{eq:MaxAeff}
\end{equation}
where $\M{F}_p\M{I}$ is proportional to the $\UV{e}$-component of the far field in the direction $\UV{k}$, and $\M{I}^{\herm}\M{F}_p^{\herm}\M{F}_p\M{I}$ to the corresponding partial radiation intensity~\cite{Gustafsson+Capek2019}. The solution of~\eqref{eq:MaxAeff} is
\begin{equation}
	\max A_{\T{eff}_p} = \lambda^2\M{F}_p^{\herm}\M{R}^{-1}\M{F}_p, 
\label{eq:MaxEff2}
\end{equation}
where the final terms in~\eqref{eq:sumefficiency} are recognized.
Inserting~\eqref{eq:MaxEff2} into~\eqref{eq:sumefficiency} produces the estimate
\begin{equation}
\sum_{n=1}^N \nu_n 
=\frac{1}{\lambda^2}\sum_{p=1}^{N_\T{R}}
\Lambda_p^2
\max\Aeff(\UV{k}_p,\UV{e}_p),
\label{eq:NDoFMaxAeffQuad}
\end{equation}
whereby refining the quadrature rule over spatial points, the sum approaches the integral~\eqref{eq:NDoFMaxAeffInt}.

\section{Mutual shadow length}\label{S:Geo2Dlines}
The mutual shadow length from a set of lines described by their end-points $\rv_{mn}$, with $m\in\{\T{T},\T{R}\}$ is determined for an illumination in the $\UV{k}$-direction, using a plane wave $E_\T{z}(\rv)=\eu^{-\ju k\UV{k}\cdot\rv}$ polarized in the $\UV{z}$-direction, see Fig.~\ref{fig:LinesShadowGeo}. The shadow is projected on a line perpendicular to $\UV{k}=\UV{x}\cos(\phi)+\UV{y}\sin(\phi)$, which using polar coordinates with azimuth angle $\phi$ results in
\begin{equation}
    \UV{p} = -\UV{x}\sin(\phi)+\UV{y}\cos(\phi).
\end{equation}
Projecting the node points on the $\UV{p}$-line produces a set of numbers $p_{mn}=\UV{p}\cdot\V{r}_{mn}$ and the lines a set of intervals $I_m=[\min_n p_{mn},\max_n p_{mn}]$. The mutual shadow is formed by $I_1\cap I_2$. The intersection is empty if $\max_n p_{mn}\leq \min_n p_{qn}$ for some $m\neq q$, \ie
\begin{equation}
    \max_n\{p_{1n}\} \leq \min_n\{p_{2n}\}
    \quad\text{or }
    \max_n\{p_{2n}\} \leq \min_n\{p_{1n}\}.
\end{equation}
The length of the interval is finally
\begin{equation}
    |\sregTR| = \max_n\{p_{1n}\} - \min_n\{p_{2n}\}
\end{equation}
or
\begin{equation}
    |\sregTR| = \max_n\{p_{2n}\} - \min_n\{p_{1n}\}.
\end{equation}
The total mutual shadow length is determined by integration~\eqref{eq:LTR}.

\section{Shadow area for two spheres}\label{S:Sphshadowarea}
Shadow area and mutual shadow area are illustrated for two spheres having radii $a_n$, $n=1,2$ and separated by a distance $h$, see Fig.~\ref{fig:ShadowAreaSph}. The spheres are, without loss of generality, placed on the z-axis, producing a BoR object with an incident illumination parametrized by the polar angle $\theta$.

The shadow of each sphere is a circular disc with the same radius as the sphere. The two shadows can overlap, partly overlap, or be separated depending on the illumination angle $\theta$ through the distance $d=h\sin\theta$. For $0\leq\theta\leq\theta_1$ with $\sin\theta_1=|a_1-a_2|/h$, the two shadows overlap, as illustrated in the bottom part of Fig.~\ref{fig:ShadowAreaSph}. For $\theta_1<\theta<\theta_2$ with $\sin\theta_2=(a_1+a_2)/h$, the two shadows partly overlap. And for $\theta_2\leq\theta\leq\pi/2$ the two shadows are separated. This behavior is repeated symmetrically for $\pi/2\leq\theta\leq\pi$. 

\begin{figure}
    \centering

    \includegraphics[width=\figw\columnwidth]{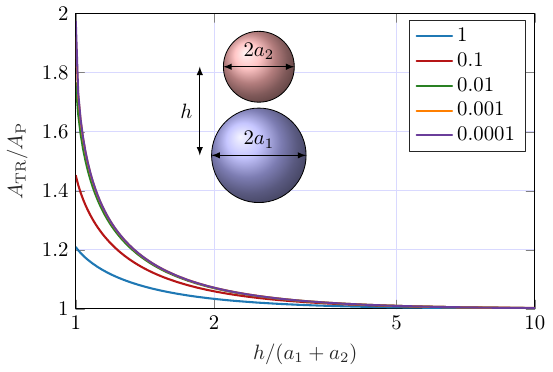}
    
    \caption{Mutual shadow area for two spheres with radii $a_1$ and $a_2$ normalized by the paraxial approximation~\eqref{eq:sphparaxial}. The quotient $a_2/a_1$ is given in the legend.}
    \label{fig:ISAsph}
\end{figure}

The total shadow area is determined by integrating over all incident angles, which reduces to a line integral over $[0,\pi/2]$ for this specific case.  For the mutual shadow area, it is sufficient to integrate over $[0,\theta_2]$. The mutual area for $\theta<\theta_1$ is $\pi a_1^2$ giving 
\begin{equation}
    A = 2\pi^2\min\{a_n^2\}(1-\cos\theta_1)
    +2\pi\int_{\theta_1}^{\theta_2} A(\theta)\sin\theta\diff\theta.
\end{equation}
The partly overlapping shadow area is divided into two parts  $A=A_1+A_2$ determined as
\begin{equation}
    A_n = a_n^2\big(\acos(d_n) - d_n\sqrt{1-d_n^2}\big)
\end{equation}
for $n=1,2$ with
\begin{equation}
    d_n = (d^2+2a_n^2-a_1^2-a_2^2)/(2da_n).
\end{equation}
The integral simplifies for two equal spheres $a_1=a_2$ and can be expressed in elliptic integrals, however it is also trivially solved by numerical quadrature, see Fig.~\ref{fig:ISAsph}.  
The mutual shadow area is normalized by the paraxial approximation
\begin{equation}
    A_\T{p} = \frac{A_1 A_2}{h^2} = \frac{\pi^2 a_1^2 a_2^2}{h^2},
    \label{eq:sphparaxial}
\end{equation}
which approximates the mutual shadow area very well for distances $h>2(a_1+a_2)$, see Fig.~\ref{fig:ISAsph}.

\end{document}